\documentclass[fleqn,usenatbib]{mnras}
\usepackage{newtxtext,newtxmath}
\usepackage[T1]{fontenc}
\usepackage{ae,aecompl}

\usepackage{graphicx, graphbox}
\usepackage{amsmath}
\usepackage{float}
\usepackage{subcaption}
\usepackage{bbm}
\usepackage{eso-pic} 
\usepackage{soul}

\AddToShipoutPictureBG*{%
  \AtPageUpperLeft{%
    \hspace{0.75\paperwidth}%
    \raisebox{-3.5\baselineskip}{%
      \makebox[0pt][l]{\textnormal{DES-2020-0542}}
}}}%

\AddToShipoutPictureBG*{%
  \AtPageUpperLeft{%
    \hspace{0.75\paperwidth}%
    \raisebox{-4.5\baselineskip}{%
      \makebox[0pt][l]{\textnormal{FERMILAB-PUB-20-653-AE}}
}}}%

\title[Joint redshift--stellar mass PDFs with RF]{A machine learning approach to galaxy properties: joint redshift--stellar mass probability distributions with Random Forest}

\author[S. Mucesh et al.]{
\parbox{\textwidth}{
\Large
S.~Mucesh,$^{1}$\thanks{E-mail: \href{mailto:sunil.mucesh.18@ucl.ac.uk}{sunil.mucesh.18@ucl.ac.uk}}
W.~G.~Hartley,$^{2}$
A.~Palmese,$^{3, 4}$
O.~Lahav,$^{1}$
L.~Whiteway,$^{1}$
A.~F.~L.~Bluck,$^{5, 6}$
A.~Alarcon,$^{7}$
A.~Amon,$^{8}$
K.~Bechtol,$^{9}$
G.~M.~Bernstein,$^{10}$
A.~Carnero~Rosell,$^{11, 12}$
M.~Carrasco~Kind,$^{13, 14}$
A.~Choi,$^{15}$
K.~Eckert,$^{10}$
S.~Everett,$^{16}$
D.~Gruen,$^{8, 17, 18}$
R.~A.~Gruendl,$^{13, 14}$
I.~Harrison,$^{19}$
E.~M.~Huff,$^{20}$
N.~Kuropatkin,$^{3}$
I.~Sevilla-Noarbe,$^{21}$
E.~Sheldon,$^{22}$
B.~Yanny,$^{3}$
M.~Aguena,$^{23, 24}$
S.~Allam,$^{3}$
D.~Bacon,$^{25}$
E.~Bertin,$^{26, 27}$
S.~Bhargava,$^{28}$
D.~Brooks,$^{1}$
J.~Carretero,$^{29}$
F.~J.~Castander,$^{30, 31}$
C.~Conselice,$^{19, 32}$
M.~Costanzi,$^{33, 34}$
M.~Crocce,$^{30, 31}$
L.~N.~da Costa,$^{24, 35}$
M.~E.~S.~Pereira,$^{36}$
J.~De~Vicente,$^{21}$
S.~Desai,$^{37}$
H.~T.~Diehl,$^{3}$
A.~Drlica-Wagner,$^{3, 4, 38}$
A.~E.~Evrard,$^{36, 39}$
I.~Ferrero,$^{40}$
B.~Flaugher,$^{3}$
P.~Fosalba,$^{30, 31}$
J.~Frieman,$^{3, 4}$
J.~Garc\'ia-Bellido,$^{41}$
E.~Gaztanaga,$^{30, 31}$
D.~W.~Gerdes,$^{36, 39}$
J.~Gschwend,$^{24, 35}$
G.~Gutierrez,$^{3}$
S.~R.~Hinton,$^{42}$
D.~L.~Hollowood,$^{16}$
K.~Honscheid,$^{15, 43}$
D.~J.~James,$^{44}$
K.~Kuehn,$^{45, 46}$
M.~Lima,$^{23, 24}$
H.~Lin,$^{3}$
M.~A.~G.~Maia,$^{24, 35}$
P.~Melchior,$^{47}$
F.~Menanteau,$^{13, 14}$
R.~Miquel,$^{29, 48}$
R.~Morgan,$^{9}$
F.~Paz-Chinch\'{o}n,$^{14, 49}$
A.~A.~Plazas,$^{47}$
E.~Sanchez,$^{21}$
V.~Scarpine,$^{3}$
M.~Schubnell,$^{36}$
S.~Serrano,$^{30, 31}$
M.~Smith,$^{50}$
E.~Suchyta,$^{51}$
G.~Tarle,$^{36}$
D.~Thomas,$^{25}$
C.~To,$^{8, 17, 18}$
T.~N.~Varga,$^{52, 53}$
and R.D.~Wilkinson$^{28}$
\begin{center} (DES Collaboration) \end{center}
\emph{\normalsize Affiliations are listed at the end of the paper}
}}

\date{Accepted 2021 January 16. Received 2021 January 15; in original form 2020 December 14}

\pubyear{2020}

\begin{document}
\label{firstpage}
\pagerange{\pageref{firstpage}--\pageref{lastpage}}
\maketitle

\renewcommand{\thefootnote}{\fnsymbol{footnote}}

\begin{abstract}
We demonstrate that highly accurate joint redshift--stellar mass probability distribution functions (PDFs) can be obtained using the Random Forest (RF) machine learning (ML) algorithm, even with few photometric bands available. As an example, we use the Dark Energy Survey (DES), combined with the COSMOS2015 catalogue for redshifts and stellar masses. We build two ML models: one containing deep photometry in the $griz$ bands, and the second reflecting the photometric scatter present in the main DES survey, with carefully constructed representative training data in each case. We validate our joint PDFs for $10,699$ test galaxies by utilizing the copula probability integral transform and the Kendall distribution function, and their univariate counterparts to validate the marginals. Benchmarked against a basic set-up of the template-fitting code \texttt{BAGPIPES}, our ML-based method outperforms template fitting on all of our predefined performance metrics. In addition to accuracy, the RF is extremely fast, able to compute joint PDFs for a million galaxies in just under $6$ min with consumer computer hardware. Such speed enables PDFs to be derived in real time within analysis codes, solving potential storage issues. As part of this work we have developed \texttt{GALPRO}\footnotemark, a highly intuitive and efficient Python package to rapidly generate multivariate PDFs on-the-fly. \texttt{GALPRO} is documented and available for researchers to use in their cosmology and galaxy evolution studies.
\end{abstract}




\section{Introduction}
\label{section:introduction}
The next generation of photometric surveys such as the Rubin Observatory Legacy Survey of Space and Time (LSST; \citealt{lsst}) and \emph{Euclid} \citep{euclid} will observe billions of galaxies.\let\thefootnote\relax\footnotetext{\textbf{\dag} \ \url{https://galpro.readthedocs.io/}} The sheer amount of data generated will enable studies ranging from the cosmic large-scale structure, to the formation and evolution of galaxies, to be conducted in unprecedented detail; ultimately leading to a transformation in our understanding of the Universe. However, one of the key challenges will be developing algorithms that can quickly and reliably extract physical properties and redshifts of galaxies.

The success of many scientific analyses critically hinges on redshift measurements. For example, redshifts are required in weak lensing tomography \citep{hu}; one of the primary probes to unveil the nature of dark energy. As a result, a large number of methods now exist to estimate redshifts from photometric data (photo-\textit{z}s) \citep[see][for a review]{salvato}. In general, they are either physically motivated or data driven. 

Template-fitting methods fall into the former category as they require prior knowledge in the form of template spectral energy distributions (SEDs). These templates are fit to the observed fluxes, and photo-\textit{z}s are usually determined using chi-square minimisation \citep[e.g.][]{hyperz}. \citet{baum} originally applied template-fitting to estimate photo-\textit{z}s of elliptical galaxies. Since then, a plethora of codes has been developed for the task such as \texttt{HYPERZ} \citep{hyperz}, \texttt{BPZ} \citep{bpz}, \texttt{LEPHARE} \citep{lephare_1999}, \texttt{ZEBRA} \citep{zebra}, \texttt{EAZY} \citep{eazy} and \texttt{BCNZ2} \citep{bcnz2}.

The fundamental principle behind data-driven methods is to learn a mapping between photometry and redshift using training data. \cite{connolly} used a polynomial function for the mapping. However, since the new millennium, machine learning (ML) methods have become popular as they are able to learn more complex mappings. Once trained, ML algorithms can make predictions on `new' galaxies. As with template-fitting, a large number of ML algorithms have been used to predict photo-\textit{z}s. These include artificial neural networks (ANN; \citealt{firth}; \citealt{collister}; \citealt{sadeh}), support-vector machines (SVM; \citealt{wadekar}), self-organizing maps (\citealt{som_way}; \citealt{som_geach}; \citealt{somz}), Gaussian process regression (\citealt{way_gpr}), genetic algorithms (\citealt{hogan}), k-nearest neighbours (kNN; \citealt{ball}), boosted decision trees (\citealt{gerdes}), random forests (RF; \citealt{carlisle_2008}; \citealt{tpz}; \citealt{rau}) and sparse Gaussian framework \citep{almosallam}. Furthermore, deep learning methods have also been implemented (\citealt{hoyle}; \citealt{disanto}; \citealt{pasquet}).

Galaxies are described by a wide range of physical properties, with stellar mass, star formation rate, age, and metallicity being among the most important. Template-fitting codes such as \texttt{FAST} \citep{fast}, \texttt{CIGALE} (\citealt{cigale_1}; \citealt{cigale_2}; \citealt{cigale_3}), \texttt{MAGPHYS} \citep{magphys}, and \texttt{BMASTELLARMASSES} \citep{antonella_2020} have been specifically designed to output these quantities. Meanwhile, the application of ML in this field has been fairly limited, but literature has now begun to emerge (\citealt{acquaviva}; \citealt{stensbo}; \citealt{bonjean}; \citealt{delli}).

While single-value (point) estimates are useful, probability distribution functions (PDFs) have become increasingly important in recent years as a full characterization of the uncertainties, beyond a point estimate and an error bar, is required for accurate analyses. This has been particularly true in the role of redshifts for weak lensing cosmology \citep[e.g.][]{bonnet}, where it has been shown that using distributions instead of point estimates can improve the accuracy of cosmological measurements (\citealt{mandelbaum}; \citealt{myers}). It is possible to extract redshift PDFs using both template-fitting and ML methods. However, ML methods have recently grown in use due to their efficiency. For example, packages such as \texttt{ArborZ} \citep{gerdes}, \texttt{TPZ} \citep{tpz}, \texttt{SOMz} \citep{somz}, \texttt{SkyNet} \citep{skynet}, and \texttt{ANNz2} \citep{sadeh} all have foundations in ML. To reach a consensus on the best algorithm in terms of PDF accuracy, \cite{schmidt} and \cite{desprez} have compared a dozen or more popular algorithms from both approaches.

The redshift and physical properties of a galaxy, measured via modelling its photometry, are correlated, and thus should be described with a multivariate distribution. The commonly used marginal distributions in redshift, stellar mass, etc., constitute a loss of information and could potentially introduce biases into a scientific analysis as a result. Consequently, a new class of template-fitting codes has come to the fore such as \texttt{BAYESed} (\citealt{han_2012, han_2014, han_2019}), \texttt{BEAGLE} \citep{beagle} and \texttt{BAGPIPES} \citep{bagpipes}. They utilise Bayesian statistical techniques such as Markov chain Monte Carlo (\citealt{goodman}; \citealt{foreman}) and nested sampling algorithms (\citealt{skilling}; \citealt{feroz_2008}, \citealt{feroz_2009, feroz_2013}) to generate multivariate posterior distributions of the most important properties. By estimating redshift and physical properties simultaneously, they allow for any uncertainties on redshift to propagate to the statistical constraints on physical properties, whilst accounting for any potential correlations \citep{beagle}. The only drawback is that it is not feasible to obtain these distributions for a large number of galaxies. For example, \texttt{BAGPIPES} takes on average a few minutes to fit each galaxy, making it prohibitively expensive to fit modern data sets where sample numbers can exceed hundreds of millions, let alone upcoming surveys where the numbers will exceed a billion. Moreover, the results of the fit to each galaxy must somehow be stored in a way that is accessible to scientific analysis routines.

Based on the speed and the competitive performance of ML algorithms when used to estimate photo-\textit{z}s, it is possible that an ML approach to the problem could be promising. With this in mind, we take a significant step towards realizing the ultimate goal of extracting full posterior distributions of galaxy properties using ML by first focusing on 2D posterior distributions of redshift and stellar mass. We choose these properties as they are two of the most important and accurate to predict (\citealt{walcher}; \citealt{conroy}). Furthermore, joint PDFs are straightforward to visualize and thus ideal for uncovering any hidden correlations or degeneracies that exist between the properties.

Joint redshift--stellar mass PDFs have many potential science applications such as determining the evolution of the stellar mass function (SMF; e.g. \citealt{papovich}; \citealt{mortlock}; \citealt{capozzi}), the cross-correlation function between galaxies and galaxy groups \citep{yang}, understanding the connection between stellar mass and dark matter in galaxy clusters \citep{antonella_2016,antonella_2020}, and the stellar-to-halo mass relation \citep[SHMR; see][for an overview]{wechsler}. However, their storage remains a potential issue. Unless there is a revolution in data storage, it will not be feasible to store a large number of multivariate PDFs. To solve this dilemma, we have developed \texttt{GALPRO}, a highly intuitive and efficient \texttt{Python} package for rapid, on-the-fly generation of n-dimensional PDFs. \texttt{GALPRO} is documented and available for fellow researchers to use in their analyses at \url{https://galpro.readthedocs.io/}.

An interesting application of \texttt{GALPRO} could be to generate joint redshift--luminosity PDFs for measurements of the Hubble constant from gravitational wave events that lack an electromagnetic counterpart (\citealt{schutz, palmese2019}; \citealt*{darksiren1}). The use of full redshift PDFs rather than point estimates is very important for standard siren measurements \citep{Palmese:2020_190814}, and the inclusion of joint redshift--luminosity PDFs allows one to correctly define the selection function of the galaxy sample at the same time.

The outline of this paper is as follows. In Sec. \ref{section:ml}, we give a brief introduction to ML, describe the RF algorithm and outline the method we use to extract point estimates, marginal and joint posterior probability distributions of redshift and stellar mass. In Sec. \ref{section:data}, we describe the pre-processing steps we perform to construct the necessary data sets. In Sec. \ref{section:models}, we describe the different RF models we train and explain the motivation behind them. We compare, discuss, and validate our results in Sec. \ref{section:results}, and place them into a familiar context via a comparison to those achieved by \texttt{BAGPIPES} in Sec. \ref{section:bagpipes}. Finally, we summarize this work in Sec. \ref{section:conclusions}.

\section{Machine learning}
\label{section:ml}
ML is a subset of artificial intelligence that focuses on the development of computer algorithms that can learn to make predictions or decisions without being explicitly programmed to do so. In general, there are three types of ML algorithms: supervised, unsupervised, and reinforcement learning. With supervised learning, the algorithm is given labelled data (i.e. the correct answers), and it learns a mapping between the input and target features. On the other hand, unsupervised learning algorithms are not given any labelled data and are left to their own accord to find structure and discover hidden patterns within data. Lastly, reinforcement learning algorithms give computers the ability to interact with a dynamic environment to achieve a predefined goal.

The application of ML in astrophysics began as early as the 1990s with the use of ANNs for star--galaxy separation \citep[e.g][]{odewahn} and morphological classification of galaxies \citep[e.g][]{storrie}. With the advent of large-scale surveys such as the Sloan Digital Sky Survey (SDSS; \citealt{sdss}) and more recently, the Dark Energy Survey (DES; \citealt{des, des_overview}; \citealt{des_book}), ML algorithms have been widely adopted to cope with the enormous influx of data and to do novel science \citep[see][for a recent review]{baron}. This trend is likely to continue with the next generation of surveys such as the Rubin Observatory LSST (\citealt{lsst}) and \emph{Euclid} \citep{euclid} as they will produce an order of magnitude more data than the previous. In the next section we describe the RF algorithm and outline our method for estimating joint redshift--stellar mass posterior probability distributions.

\subsection{Random forest}
\label{section:random_forest}
RF (\citealt{random_forest}) is a supervised learning algorithm based on ensemble learning, as it utilizes many decision trees to make predictions. These trees are a type of data structure, which allow one to make a decision using a series of yes/no questions, and they can be used for regression and classification. They are built using a recursive algorithm that splits the data usually into two groups at each step until some predefined threshold is achieved. The job of the algorithm is to identify groups that have similar input and target features and is therefore related to the kNN algorithm \citep{knn}. The main components of the decision tree are the root, decision, and leaf nodes. The root node defines the first and the most optimum split. The decision nodes describe the subsequent splits, and the leaf nodes contain the final groups.

The exact process of building a decision tree is as follows. At each step, all possible splits are evaluated in all dimensions of the input feature space. For classification, the data are split to best separate different classes, and this is achieved by maximizing the information gain, or in other words, minimizing the impurity using metrics such as the information entropy, Gini entropy, and classification error \citep{tpz}. For regression, the data are split such that the average values of the target variable are representative of the groups. Usually, the variance is minimized to accomplish this using the loss function:

\begin{equation}
\label{eq:1}
    S = \frac{1}{n_{m}} \sum_{m} \sum_{i \epsilon m} (\Tilde{y}_{i} - \bar{y}_{m})^2,
\end{equation}

\noindent
where $n_{m}$ is the number of data points in a group, $m$, $\Tilde{y}_{i}$ are the values of the target variable in the group, and $\bar{y}_{m}$ is the group mean of the target variable.

Once the decision tree is built or trained, it can be used to make predictions. If the training data used to build the tree are complete and representative, then a new datum will end up in a leaf node that is representative of itself. The content of the leaf node can then be used to make a prediction. For classification, the prediction is the mode, and for regression, it is the mean of the leaf node.

The simplicity of the decision tree algorithm makes it one of the most popular learning mechanisms. However, decision trees only perform well on training data as they are prone to overfitting. The RF algorithm solves this issue by constructing multiple decision trees and by making a few adjustments. For example, when building the decision trees, only a subset of the training data and features is used. This technique is called feature bagging and injects randomness, making RFs more flexible and better suited to make predictions on data not encountered before. By using multiple decision trees in combination with feature bagging, RF aims to preserve the low bias of a single decision tree whilst simultaneously reducing variance to successfully navigate the bias-variance tradeoff. In summary, a RF can be built using the following process:

\begin{enumerate}
    \item Create a bootstrapped data set by sampling randomly from the training data with replacement.
    \item Choose a random subset of input features when building a decision tree using the bootstrapped data.
    \item Repeat the process to build multiple decision trees, thus creating a `forest'.
\end{enumerate}

The process of predicting with a RF is similar to predicting with a single decision tree. The only difference is that predictions from all the decision trees are collated. For classification, the prediction is the most predicted class, and for regression, it is the mean of all the values predicted by the decision trees. As is the case with many ML algorithms, RF has hyperparameters, which need to be specified beforehand. These hyperparameters can be tuned to give the best performance, and some of the most important are as follows:

\begin{itemize}
    \item \texttt{n\_estimators} -- The number of decision trees used to build the RF determines its effectiveness. Each decision tree is built using a subset of training data. As a result, if the number used is too small, then the likelihood of complete coverage of the training data decreases, resulting in poor performance. The performance improves as the number of trees increases, but at a cost, which is the time taken for training. The key is to find the right balance between performance and training time because the gains become negligible after a certain point.
    \item \texttt{max\_features} -- The maximum number of features considered at each step when building the decision trees controls the correlation between them and hence, the flexibility of the RF. Usually, $\sqrt{N}$ features are sufficient to build each decision tree, where $N$ is the total number of input features. 
    \item \texttt{max\_depth} -- The maximum depth defines the number of levels in the decision tree, and it determines how finely or coarsely the training data are grouped. A low depth leads to underfitting, and if the depth is too high, it may lead to overfitting. In essence, the maximum depth provides a stopping criterion. The minimum number of training samples in a leaf node (\texttt{min\_samples\_leaf}), and the minimum number of training samples in a leaf node before the data are split (\texttt{min\_samples\_split}) also serve the same purpose. 

\end{itemize}

\subsection{Joint PDF estimation method}
\label{section:method}
The RF algorithm has previously been utilized to extract point estimates \citep{carlisle_2008, carlisle_2010} and PDFs \citep{tpz} of redshift. Recently, \cite{bonjean} used the algorithm to predict stellar masses and star formation rates of galaxies. They built a single model to predict both target variables simultaneously. The process of building decision trees to achieve this is conceptually similar to building them to predict one target variable. The only difference is that at each step, to decide the best split, the average loss function for two or more variables is minimized. In eq. \ref{eq:1}, $\Tilde{y}_{i}$ and $\bar{y}_{m}$ are now a vector of target variables and the means, respectively. As this loss function is scale dependant, the target variables must be transformed to place them on scales with similar ranges otherwise the variance of one will dominate, resulting in the algorithm expending more effort in getting one target variable correct at the expense of others \citep{breskvar}. Once trained, the leaf nodes in the decision trees contain values of the target variables.

We apply this methodology to predict redshift and stellar mass simultaneously, thus preserving any correlation between the properties. As both variables are continuous, we use regression trees to build the forest. However, it is entirely possible to use classification trees as shown by \cite{gerdes} and \cite{tpz}. Another motivation for using regression trees is that they are generally faster to train and better suited to non-uniform data. To summarize the process,

\begin{itemize}
    \item Galaxies cluster together in n-dimensional space if they have comparable values of input features.
    \item The algorithm identifies these clusters by minimizing the loss function (eq. \ref{eq:1}), with redshift and stellar mass being the target variables.
    \item These clusters end up in the leaf nodes of the decision trees. In the end, the leaf nodes contain redshifts and stellar masses of similar galaxies.
\end{itemize}

We extract point estimates of redshift and stellar mass by running a `new' galaxy down all the decision trees and using the mean of all the predicted values. To build marginal posterior distributions, we aggregate the values of redshift and stellar mass in the leaf nodes across all the decision trees, respectively. Finally, we combine the aggregated values to build joint posterior distributions. We would like to point out that our method is flexible and can be adapted to generate joint PDFs of any other combination of properties. However, we chose redshift and stellar mass as they are two of the most important and accurate properties to predict. Furthermore, the method is flexible and can be applied to generate n-dimensional PDFs. We describe the implementation of the RF in this work, and the input features in Sec. \ref{section:models}.

\section{Data}
\label{section:data}
We use data from two different surveys to train and test our RF models. These are the DES (\citealt{des, des_overview}; \citealt{des_book}) and the Cosmological Evolution Survey (COSMOS; \citealt{cosmos}).

\subsection{Cosmological Evolution Survey}
\label{subsection:COSMOS}
The COSMOS observed a $2 \deg^{2}$ equatorial field in the entire spectral range from radio to X-ray with both ground and space-based telescopes, collecting photometric and spectroscopic data. In this field, $\sim 2$ million galaxies were detected, spanning $75\%$ of the age of the Universe \citep{cosmos}.

We use the COSMOS2015 \citep{cosmos2015} catalogue from the field for its photo-\textit{z}s and stellar masses. Usually, to train an ML algorithm to predict photo-\textit{z}s, spectroscopic redshifts (spec-\textit{z}s) are used. However, the photo-\textit{z}s in this catalogue have been shown to be precise and accurate. Compared to photo-\textit{z}s from surveys such as the DES and the SDSS (\citealt{sdss}), the COSMOS photo-\textit{z}s have been computed using more than $30$ bands spanning a huge portion of the electromagnetic spectrum, as opposed to four or five optical bands. The most precise photo-\textit{z}s have been estimated for very bright, low redshift, star-forming galaxies, with a normalized median absolute deviation (NMAD; \citealt{nmad}) of $0.007$, of which $0.5 \%$ are catastrophic outliers. Furthermore, in the deepest regions of the survey, $90 \%$ of galaxies with stellar mass greater than $10^{10} M_{\odot}$ at $z = 4$ have been detected. The high photo-\textit{z} precision and the overall completeness of the survey in stellar mass makes this an exemplary data set to use in this work.

\subsection{Dark Energy Survey}
\label{subsection:DES}
The DES is a visible and near-infrared survey that has imaged $\sim 5100 \deg^{2}$ of the South Galactic Cap $10$ times in $grizY$ photometric bands using the Dark Energy Camera (DECam; \citealt{decam}) over a span of $6$ yr, starting in $2013$. It is expected to have generated $\sim 310$ million galaxies with photo-\textit{z}s, once all the data has been processed. In addition, the survey targeted a set of four fields with a total of $10$ DECam pointings over $27 \deg^{2}$ for supernova (SN) science. This SN survey had an approximately weekly cadence and thus many more epochs per pointing than the main survey (\citealt{neilsen}). We use two data sets from the DES survey, which are discussed in the following sections.

\subsubsection{DES Deep Fields}
As part of the DES Year 3 (Y3) cosmology analysis, observations from the SN survey were combined with community data, additional DES exposures (particularly in $u$-band) and coincident near-infrared data to form the DES Deep Fields (DF) catalogue (\citealt*{deep_fields}). The principal aims of the DF project are to improve calibration of redshift distributions in the main survey and to act as a prior on the population of full multicolour images for \texttt{Balrog} (discussed in the next section), to better understand the systematics and selection function of the wide-field (WF) survey. These goals rely on the fact that the DF represents a statistically complete, yet effectively noiseless, population of the galaxies that are found in the WF survey. Other motives include conducting galaxy evolution studies, science with the faintest possible sources and the properties of the host galaxies of transient events. 

The Y3 DF catalogue consists of data from three SN fields plus the COSMOS field, with a total coverage of $5.88 \deg^{2}$ and photometry of over $1.7$ million objects (after masking for image defects) in DECam $ugriz$ and VIRCam $JHK$ bands. We combine the deep ($\sim 1.25$ mag fainter than the WF data) and precise $griz$ photometry in this catalogue with the accurate redshifts and stellar masses from the COSMOS2015 catalogue to produce a `baseline' DF data set. Specifically, we utilise the bulge+disc model-fit magnitudes computed using the Multi-Object Fitting (\citealt{mof}) algorithm.

Our goal is to produce valid posterior PDFs of galaxies in the main DES survey and to achieve this we require a suitable data set with which to train a RF model. The photometric errors in the DF data set would not reflect those in the WF and so would lead to biased results if used directly as training data. Furthermore, the COSMOS field does not overlap the main survey area and the redshifts and stellar masses that could be derived from model fitting to the four-band WF data are grossly imprecise compared to those in the COSMOS2015 catalogue. In essence, we require a catalogue of DF galaxies that emulate galaxies in the WF to overcome these issues, and for this, we take advantage of the \texttt{Balrog} algorithm.

\subsubsection{Balrog}
\label{section:balrog}
\texttt{Balrog} is a \texttt{Python} package designed for the purpose of measuring the selection function of imaging surveys (\citealt{balrog_1}; \citealt{balrog_2}). The process by which it achieves this is as follows. A realistic ensemble of fake stars and galaxies are generated using \texttt{GALSIM} \citep{galsim}, including survey characteristics appropriate to their intended sky location, e.g. seeing FWHM. The fake objects are then embedded into real survey images, thus inheriting many of their properties. Finally, the objects are detected and measured using \texttt{SExtractor} \citep{sextractor} in the same way as the original survey images. The output catalogue comprises a Monte Carlo sampling of the selection function and measurement biases and naturally accounts for systematic effects arising from the photometric pipeline, detector defects, seeing and other sources of observational systematic errors.

\begin{figure*}
    \centering
    \includegraphics[width=0.96\textwidth]{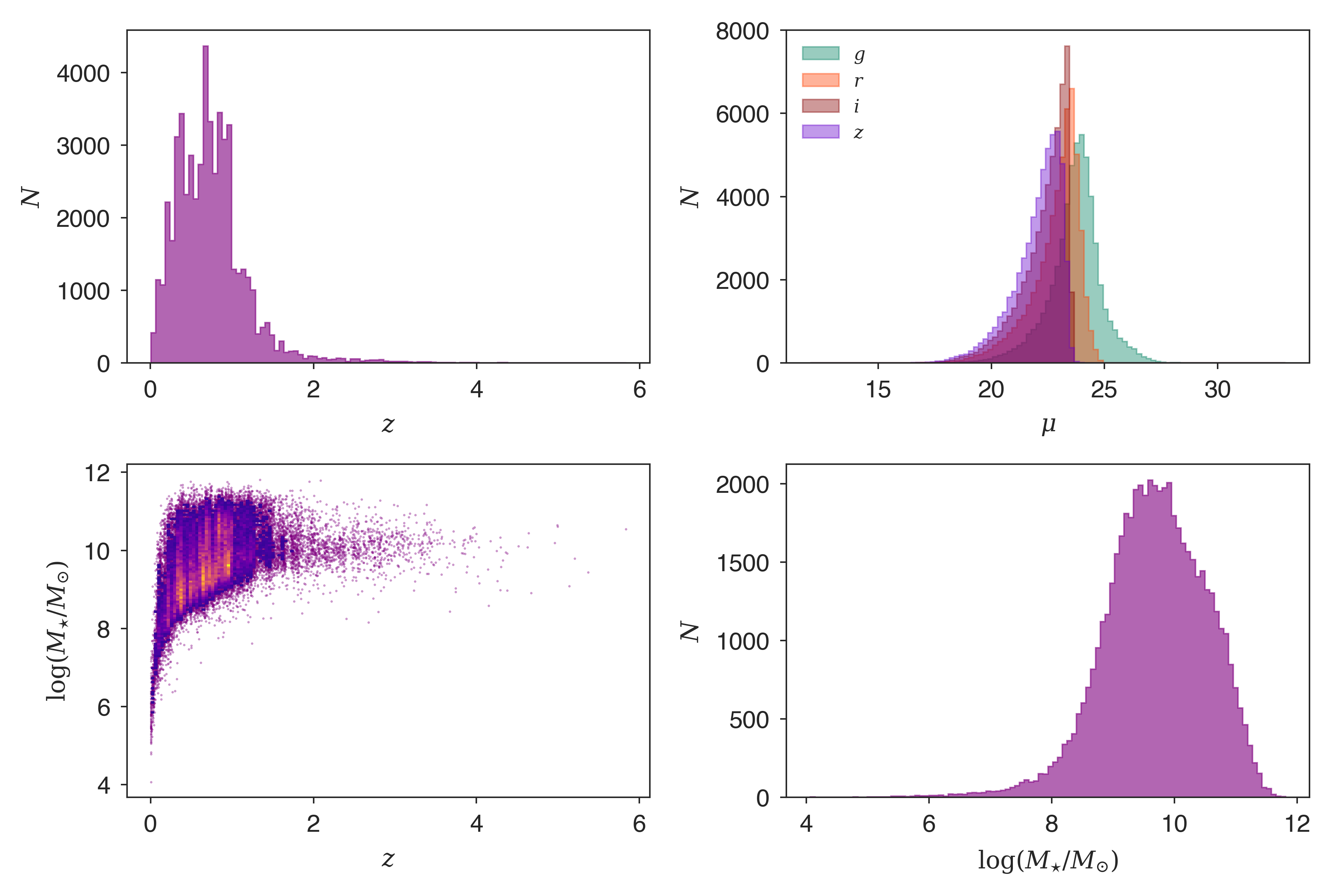}
    \caption{Marginal and joint distributions of redshifts and stellar masses of galaxies in the DF data set and the distributions of $griz$ luptitudes. The colours in the joint distribution indicate the density of points. The DF data set is created by cross-matching galaxies in the DES Y3 Deep Fields (DF) and the COSMOS2015 catalogues. All galaxies with erroneous redshift and stellar masses are discarded from the data set, and a magnitude-limited sample is produced by selecting galaxies with $i<23.5$. The $griz$ luptitudes in the data set are computed from fluxes in the Y3 DF catalogue, while the redshifts and stellar masses are from the COSMOS2015 catalogue.}
\label{fig:distributions}
\end{figure*}

The Balrog process requires a prior population of galaxies from which to draw objects. The DES Y3 Balrog catalogue (\citealt{balrog_2}) was produced by injecting model fits of galaxies drawn randomly from the Y3 DF catalogue into DES Y3 single-epoch images and then measuring their properties. This catalogue contains true and measured $griz$ photometry of nearly $4$ million objects, and it provides us with ready-made emulated galaxies that reflect our target WF data set, the DES Y3 GOLD (\citealt{sevilla}). By combining the Y3 Balrog catalogue with COSMOS2015, we obtain a data set that closely matches and is representative of the WF data, capturing many of the details of the objects' noise properties, but with the addition of accurate redshifts and stellar masses.  From the catalogue, we use composite model magnitudes in this work. In the next section we outline the pre-processing steps we perform to create the DF and WF data sets.

\subsection{Pre-processing}
\label{section:preprocessing}
To construct the DF data set, we first cross-match galaxies in the Y3 DF and the COSMOS2015 catalogues using \texttt{TOPCAT} \citep{topcat} with a matching radius of $1$ arcsec, and this serves the dual purpose of enabling the use of accurate photo-\textit{z}s (\texttt{PHOTOZ}) and stellar masses (\texttt{MASS\_BEST}) in our analysis and removing galaxies in all the other fields besides the COSMOS field. Next, we discard stars and any galaxies with erroneous redshifts and stellar masses by ensuring $0<z<9.99$, and we produce a magnitude-limited sample by selecting galaxies with $i <23.5$. These cuts automatically remove saturated objects and bad areas. We discover that there are some faint galaxies with close to zero or even negative fluxes in the $grz$ bands, resulting in their magnitudes being undefined. To solve this issue, we convert all galaxy fluxes into `asinh' magnitudes or `luptitudes' \citep{lupton}, defined as

\begin{equation}
    \mu = \mu_{0} - a \sinh^{-1}{\left(\frac{f}{2b}\right)},
\end{equation}

\noindent
where $\mu_{0} = m_{0} - 2.5\log{b}$, $a=2.5\log{e}$, $f$ is the flux, $b$ is an arbitrary softening parameter, and $m_{0}$ is the magnitude zero point. The authors state that the optimal value of $b=\sqrt{a}\sigma$, where $\sigma$ is the standard deviation of the flux. We set the value of $\sigma$ to be the median of the standard deviations. Additionally, we transform flux errors into luptitude errors using

\begin{equation}
    \sigma_{\mu} = \frac{a\sigma}{2b}. 
\end{equation}

Luptitudes behave like magnitudes for bright photometry and like fluxes for faint photometry, with the turning point in the behaviour determined by the softening parameter. By converting to luptitudes, we avoid introducing an additional selection effect by not discarding galaxies with negative fluxes.

To produce the WF data set, we start anew and match `WF' galaxies in the Y3 Balrog catalogue to their counterparts in the Y3 DF using the \texttt{ID} column. Next, we cross-match the galaxies in the intermediate catalogue to the COSMOS2015 catalogue. There are multiple scattered WF copies of each DF galaxy in the Balrog catalogue to efficiently sample the DES selection function, and to preserve this we keep all of the copies. This is an important aspect of our set-up, as it captures the selection function through the galaxy detection probability as a function of true photometry and light profile, and the asymmetric scatter between photometry and galaxy properties (redshift and stellar mass) that it induces. We remove any galaxies with erroneous flux measurements by selecting all galaxies with \texttt{MEAS\_CM\_FLAG} $=0$  (\citealt{balrog_2}). Finally, we repeat all the aforementioned cuts and steps used in constructing the DF data set, the only difference being that on this occasion, we apply the $i$-band cut to the magnitudes of WF galaxies. Thus, we have `augmented' a completely realistic target data set which effectively replicates the systematics in the WF survey without compromising on the accuracy of redshifts and stellar masses. 

After all the pre-processing steps, there are $53,491$ galaxies in the DF data set and $393,276$ galaxies in the WF data set. Each data set contains the following information: $griz$ luptitudes and luptitude errors, photo-\textit{z}s and stellar masses. Additionally, we compute all the relevant lupticolours; and the associated errors using the standard error propagation formula:
\begin{equation}
    \sigma_{c} = \sqrt{\sigma_{\mu_{1}}^{2} + \sigma_{\mu_{2}}^{2}},
    \label{eq:error_propagation}
\end{equation}

\noindent
where $\sigma_{\mu_{1}}$ and $\sigma_{\mu_{2}}$ are the errors on the luptitudes, and $\sigma_{c}$ is the error on the computed lupticolour. Fig. \ref{fig:distributions} shows the marginal and the joint distribution of redshifts and stellar masses of galaxies in the DF data set, and the distributions of $griz$ luptitudes. The average redshift and stellar mass is approximately $0.7$ and $5 \times 10^{9} M_\odot$, respectively. For the sake of brevity, we do not show a similar figure for the WF data set as the distributions are broadly similar.

We perform an $80:20$ split on the DF and WF data sets to create their training and testing data sets, respectively. As there are multiple copies of each galaxy in the WF data set, we ensure that there is no admixture of unique galaxies in its training and testing data sets. In other words, unique galaxies that exist in the training data set do not appear in the testing data set, and vice versa. As a consequence, there are $314,196$ and $79,080$ galaxies in the WF training and testing data sets, respectively. Lastly, we randomly sample $10,699$ galaxies without replacement from the WF testing data set to construct its final version. We do this to ensure that the number of galaxies in both the DF and WF testing data sets matches, thus enabling us to make a fair comparison when testing our RF models. 

The training data sets represent prior information that the RF models utilize in order to make predictions on the test data sets. As a result, one must construct a suitable and representative training data set (as we have done) when using outputs from an ML model in their scientific analysis. In the next section we describe the different RF models, explain the motivation behind them, and the implementation of the RF algorithm we use in this work.

\section{Models and Implementation}
\label{section:models}
We train and test two different RF models, with redshift and stellar mass as the target variables and the following as input features:

\begin{itemize}
    \item $griz$ luptitudes
    \item $griz$ luptitude errors
    \item $g-r$, $r-i$, and $i-z$ lupticolours, and their associated errors
\end{itemize}

We build the first model using the DF data set and refer to it as DES-DF from here onwards. The high-precision photometry of DF galaxies combined with the accurate redshifts and stellar masses allows us to establish the baseline performance. We build the second model to produce valid posterior PDFs of galaxies in our target data set (the DES Y3 GOLD) by training on the WF data set. We refer to this model as DES-WF. 

To train and test our RF models, we use the implementation of the algorithm in the \texttt{Python} ML package \texttt{scikit-learn}. In particular, we use the \texttt{RandomForestRegressor} module from the package, which allows us to do regression. Before training, we do not perform feature scaling as the RF algorithm is invariant under monotonic transformations. Furthermore, we do not scale the target features because redshift and stellar mass (in the logarithmic form) have similar ranges. Besides, \texttt{scikit-learn} automatically normalizes the variances of individual target variables so that they contribute equally to the loss function. 

As previously discussed in Sec. \ref{section:random_forest}, RF has hyperparameters that can be tuned to increase the performance of a model. Therefore, we tune our RF models before training using a combination of random search and grid search. We first set-up a wide grid of hyperparameters and run the models using 100 different combinations. Next, we use a grid search around the best hyperparameters found in the previous searches. After tuning, we find that the performance of the models, in terms of the root-mean-square error (RMSE), only improves by $1-2 \%$. In principle, one could use metrics associated with the validity of PDFs (described in Secs. \ref{subsection:marginal_pdfs_validation} and \ref{subsection:joint_pdfs_validation}). However, we opted for the simple RMSE as we do not believe that there exists a single metric that can fully characterize the performance of a model. Given the insignificant improvements in the performance of our models, we ultimately resorted to using the following default \texttt{scikit-learn} hyperparameters for training both models:

\begin{itemize}
    \item \texttt{n\_estimators: 100}
    \item \texttt{max\_features: auto}
    \item \texttt{max\_depth: none}
    \item \texttt{min\_samples\_leaf: 1}
    \item \texttt{min\_samples\_split: 2}
    \item \texttt{max\_leaf\_nodes: none}
    \item \texttt{min\_impurity\_decrease: 0.0}
    \item \texttt{min\_impurity\_split: none}
    \item \texttt{min\_weight\_fraction\_leaf: 0.0}
\end{itemize}

\begin{figure*}
    \begin{subfigure}[h!]{0.48\textwidth}
    \includegraphics[width=\textwidth]{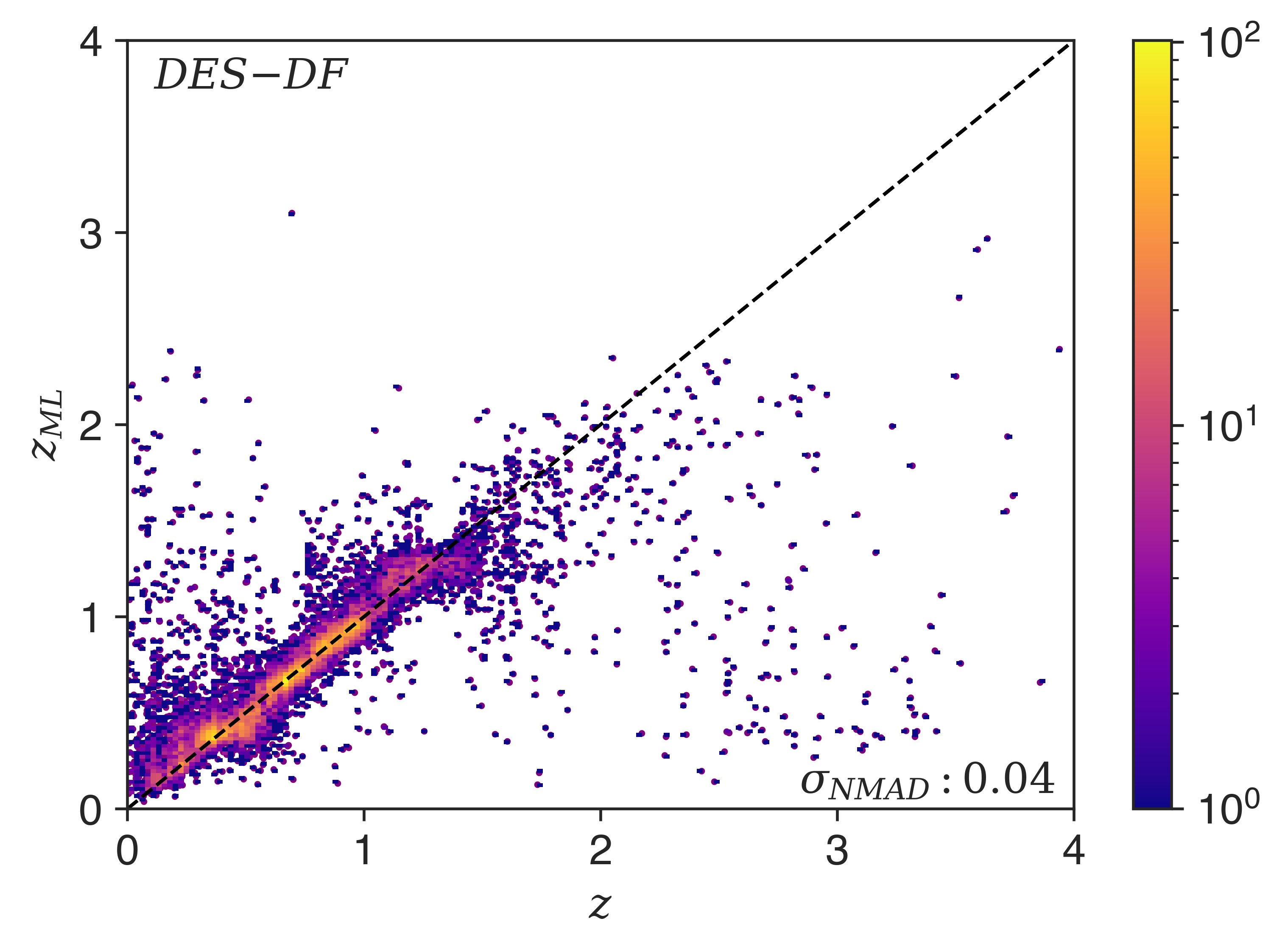}
    \end{subfigure}
   \begin{subfigure}[h!]{0.48\textwidth}
    \includegraphics[width=\textwidth]{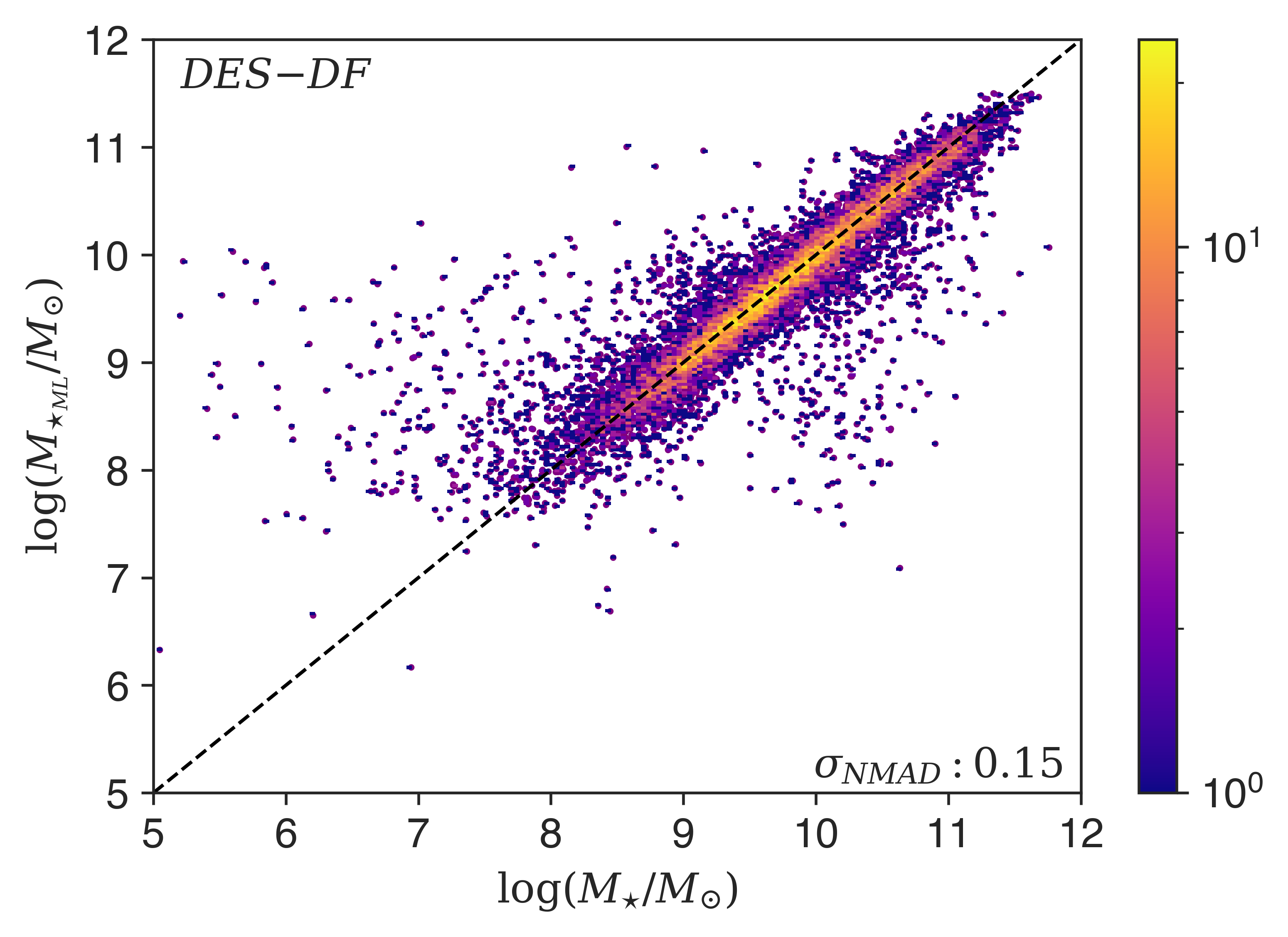}
    \end{subfigure}
    \begin{subfigure}[h!]{0.48\textwidth}
    \includegraphics[width=\textwidth]{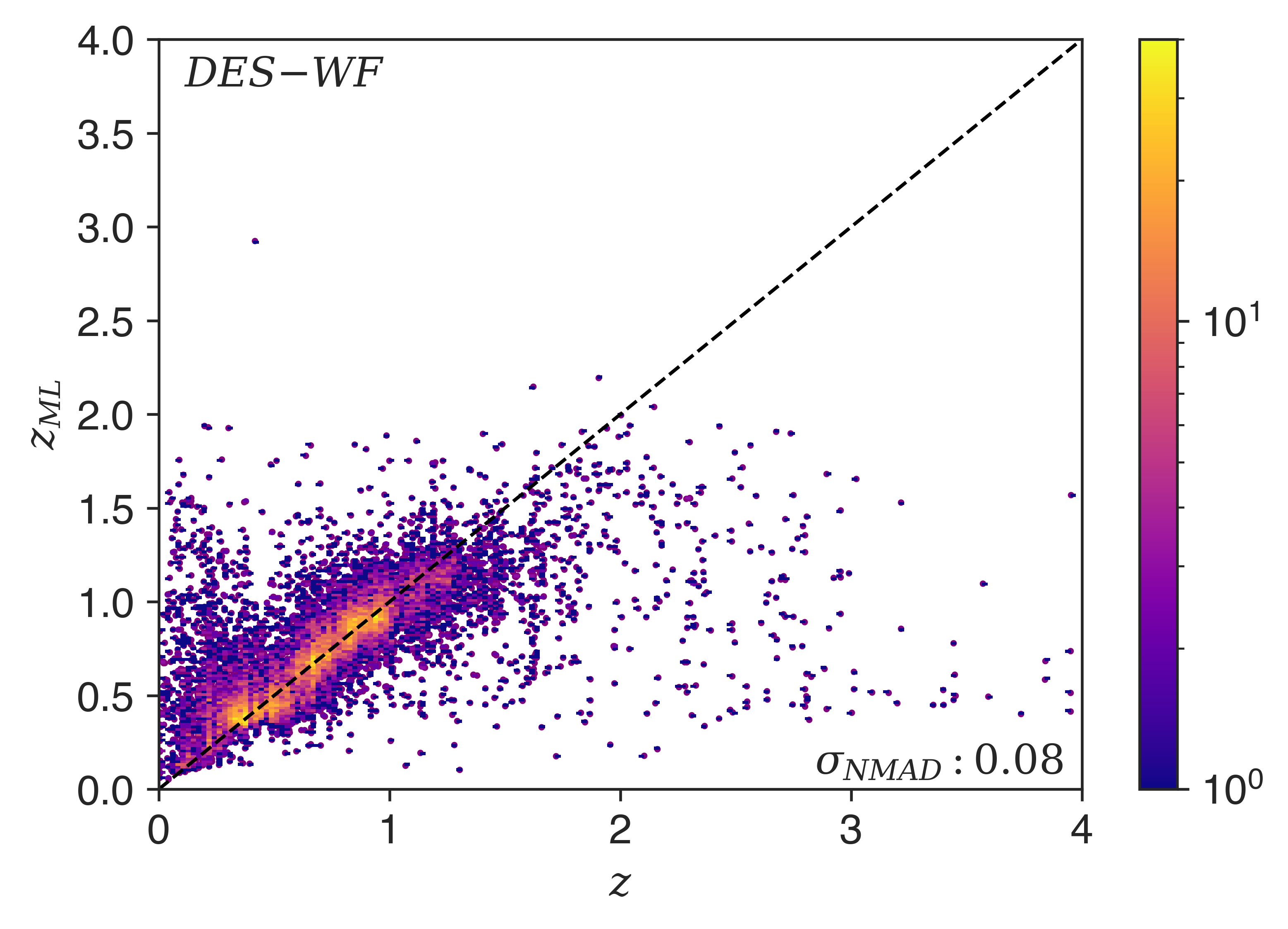}
    \end{subfigure}
   \begin{subfigure}[h!]{0.48\textwidth}
    \includegraphics[width=\textwidth]{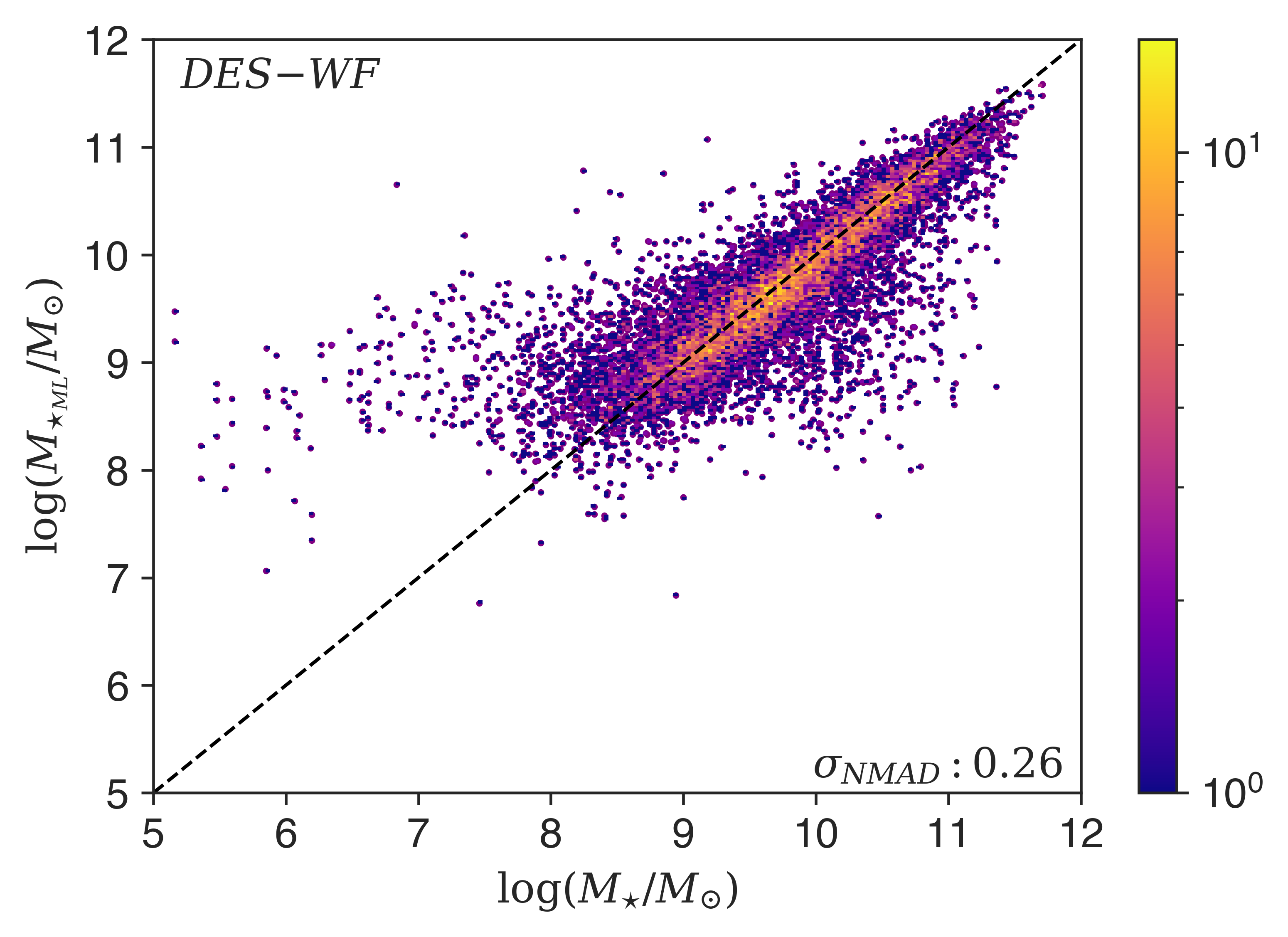}
    \end{subfigure}
\caption{`True' redshifts and stellar masses of test galaxies versus the predictions made by the DES-DF and DES-WF models. The colours indicate the density of points. The normalized median absolute deviation (NMAD; \citealt{nmad}) metric values are stated for redshift and stellar mass respectively. For redshift, the bias $\hat{y} - \Tilde{y}$ is divided by $1 + \Tilde{y}$ in Eq. \ref{eq:nmad}.}
\label{fig:scatter}
\end{figure*}

With these hyperparameters, the decision trees are fully grown, until the training data can no longer be split. We set \texttt{max\_features} to \texttt{auto} instead of $\sqrt{N}$, where $N$ is the total number of input features, to ensure that our models have sufficient prior information as we are using a limited number of photometric bands to begin with. We train and test both models on a 13" Macbook Pro (2.4 GHz Intel Core i5, 16GB LPDDR3) using \texttt{GALPRO}, and it takes less than $1$ and $5$ min, respectively, to generate PDFs for $10,699$ galaxies. In the next section, we compare, discuss, and validate the point estimates, marginal and joint posterior PDFs of redshift and stellar mass of test galaxies estimated from the trained models. 

\section{Results and Discussion}
\label{section:results}

\subsection{Point estimates}
\label{subsection:point_estimates}
We extract point estimates by averaging predictions from all the decision trees in a given RF model. In order to quantify how the models are performing, we use the NMAD metric for redshift and stellar mass. The NMAD is defined as

\begin{equation}
    \sigma_{\textrm{NMAD}} = 1.4826 \times \textrm{median} \mid \hat{y}_{i} - \Tilde{y}_{i} \mid,
\label{eq:nmad}
\end{equation}

\noindent
where $\hat{y}_{i}$ and $\Tilde{y}_{i}$ are the predicted and `true' values of redshift and stellar mass of galaxies, respectively. For redshift, the bias $\hat{y} - \Tilde{y}$ is divided by $1 + \Tilde{y}$.

Fig. \ref{fig:scatter} shows the redshifts and stellar masses of test galaxies versus the predictions made by DES-DF and DES-WF. Most of the data points lie close to the diagonal, which indicates that the predicted redshifts and stellar masses are accurate. However, there are outliers at low and high redshifts and low stellar masses. There is a lack of training data available in these regions, as can be observed in Fig. \ref{fig:distributions}. Given the strong correlation between the accuracy of a RF model and the abundance of training data, these outliers are to be expected.

Moreover, the degradation in performance could be due to degeneracies that exist in the colour--redshift space. For example, at $z<0.2$, there is a lack of strong spectral features that can be detected in the $griz$ bands. Using the $u$ band can break the degeneracies. However, we do not use it as an input feature as the band is not available in the DES data. Furthermore, in the redshift range, $1.2 < z < 2.2$, there is a lack of strong spectral features in the visible bands \citep{hyperz}. These degeneracies can lead to incorrect clustering of training galaxies and thus inaccurate point predictions. 

Comparing the two models, the point-estimate performance of DES-DF is better than DES-WF, with $\sigma_{\textrm{NMAD}}$ of $0.04$ and $0.15$ dex for redshift and stellar mass, respectively. There is a visible increase in the scatter in the DES-WF plots, and this is reflected in the values of the performance metric doubling for redshift to $0.08$ and increasing by $\sim 73\%$ to $0.26$ dex for stellar mass. This drop in performance is primarily due to the degraded photometric precision, which makes it difficult for the RF to cluster galaxies, resulting in inaccurate predictions. Nevertheless, DES-WF still performs well for a significant portion of test galaxies as can be observed. On a related note, we also explored the impact on the performance when predicting one vs two variables. We built two models each using the DF and WF data sets to predict redshift and stellar mass separately and found that there was an insignificant improvement in the performance, with $\sigma_{\textrm{NMAD}}$ decreasing by $0.001-0.002$.

\subsection{Marginal probability distributions}
\label{subsection:marginal_pdfs}
The point estimates we extracted are not perfect. In general, inaccuracies can arise from

\begin{itemize}
    \item Incomplete and incorrect information -- The information provided to an ML algorithm may not be sufficient to learn the perfect mapping between the input features and target variables. For example, to estimate redshifts to a high degree of accuracy, spectroscopic data are required. However, we use photometric data that only provides a rough sampling of the underlying SED. Furthermore, the data used for training and testing have to be accurate. In our case, the redshifts and stellar masses we use to train our RF models may contain some errors. They have been estimated using the template-fitting code \texttt{LEPHARE}, which utilizes template SEDs and they may not be a perfect representation of the true SED. Therefore, the mappings learnt by the RFs may not be entirely accurate, and this could lead to the observed errors in the estimates. Furthermore, we predict redshifts and stellar masses using four-band photometry while those in the COSMOS2015 catalogue are computed using more than 30 bands. Consequently, there will be subtle differences between our predictions and the `truth'.
    \item Unrepresentative and incomplete training data -- The lack of representative and complete training data can also lead to errors. In our case, the training data are highly likely to be representative. However, in some regions, the data are sparse, and therefore do not provide a complete sampling of the target population. For example, at low and high redshifts, the number of galaxies available for training reduces dramatically as can be observed in Fig. \ref{fig:distributions}, and this causes the performance of the algorithm to suffer. Furthermore, the effect of sample variance from the small COSMOS area can lead to some incompleteness. 
    \item ML algorithms and hyperparameters -- Different ML algorithms learn using different methods. As a result, predictions on the same datum can be slightly different. Furthermore, the hyperparameters can also have an effect, as discussed in Sec. \ref{section:random_forest}. However, the performance of ML algorithms suitable for a specific problem generally converges given sufficient and good quality training data.
\end{itemize}

In order to characterize uncertainties associated with our point estimates, we extract marginal posterior distributions of redshift and stellar mass. We do this by aggregating the redshift and stellar mass values in the leaf nodes of the decision trees in a RF that are representative of the test galaxy in question. We extract the distributions from the trained models and validate them using several techniques and metrics described in the next section.

\subsubsection{Marginal PDFs validation}
\label{subsection:marginal_pdfs_validation}

Unlike point estimates, it is not possible to validate individual redshift and stellar mass PDFs as the true distributions are not available. Consequently, we aim to determine the validity of the marginal PDFs as a whole. We use the framework developed by \cite{gneiting_2007}, which is founded on the paradigm of maximizing the sharpness of the predictive distributions subject to calibration. Sharpness refers to the concentration of predictive distributions and is a property of the distributions only. The authors describe calibration as the statistical consistency between the distributions and the truth. We refer to this as validation as it better captures the essence of use in our context. However, for consistency, we will use the former when describing the authors' work. In this paper, we focus on calibration to validate the marginal PDFs produced by our models, rather than sharpness, as the latter is useful when ranking competing calibrated methods. Furthermore, as demonstrated by \cite{bordoloi}, one could use the framework to empirically recalibrate marginal PDFs. However, this can be challenging and could potentially result in unforeseen issues. 

\begin{figure*}
  \begin{subfigure}[h!]{0.48\textwidth}
    \includegraphics[width=\textwidth]{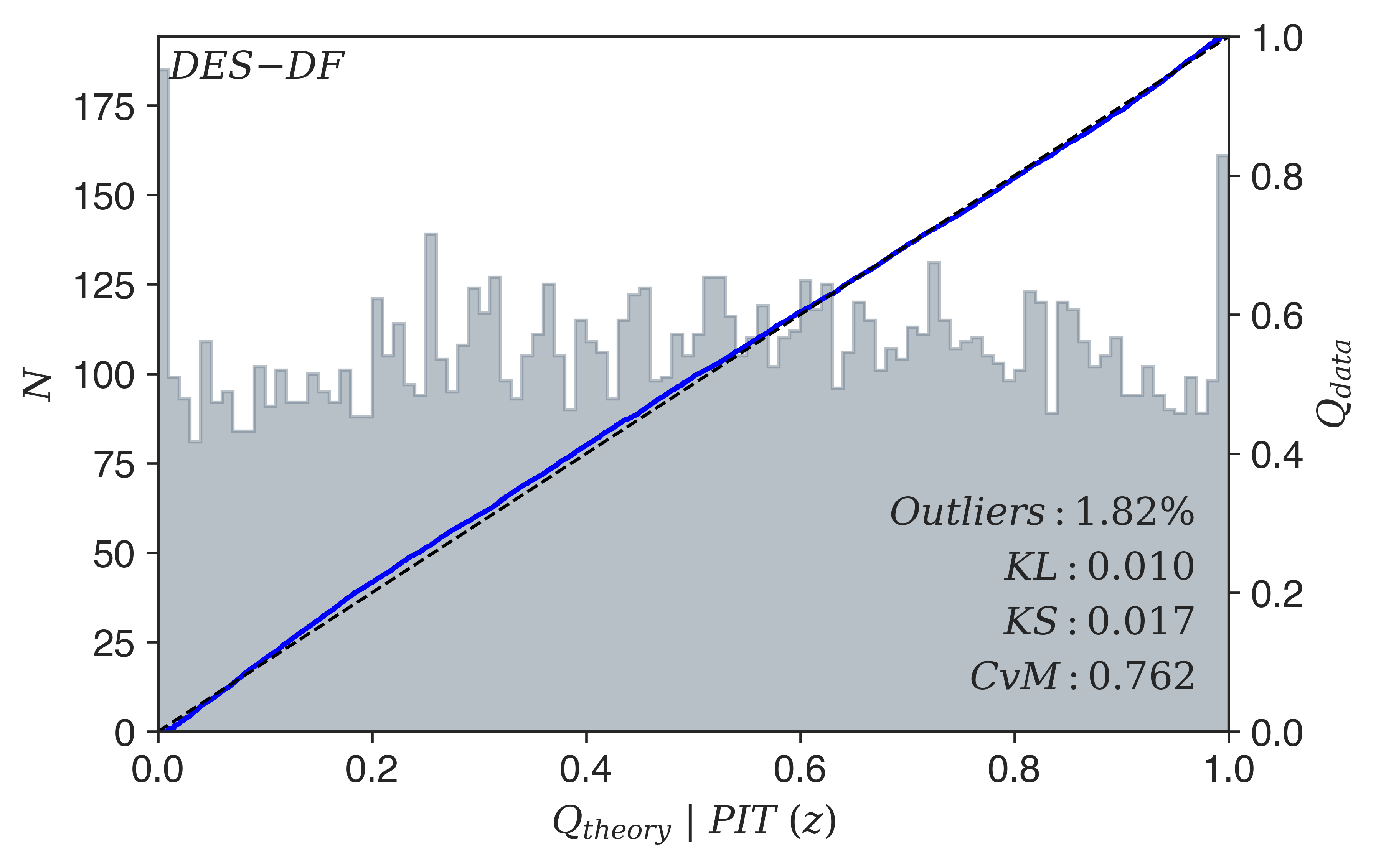}
  \end{subfigure}
  \begin{subfigure}[h!]{0.48\textwidth}
    \includegraphics[width=\textwidth]{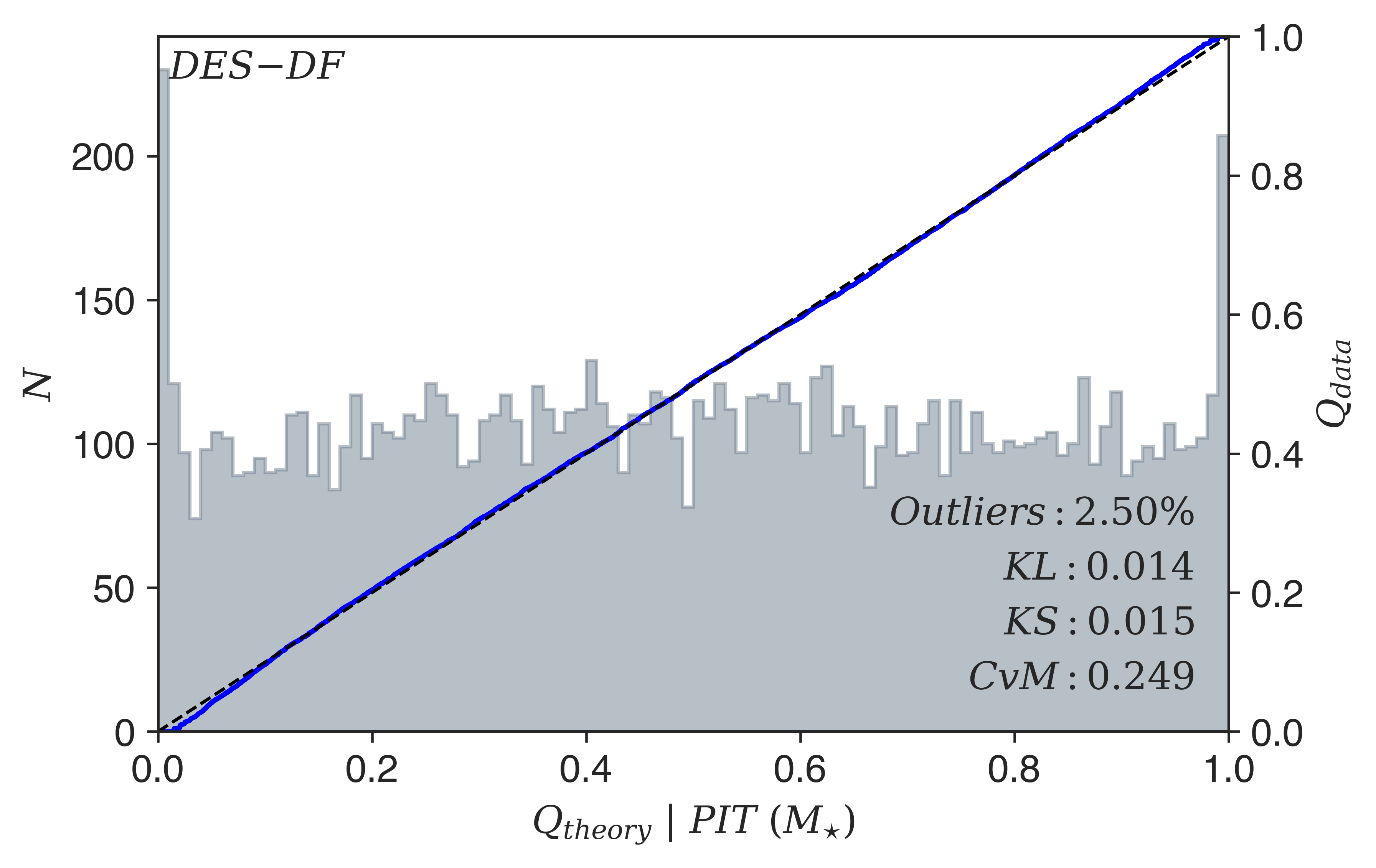}
  \end{subfigure}
 \begin{subfigure}[h!]{0.48\textwidth}
    \includegraphics[width=\textwidth]{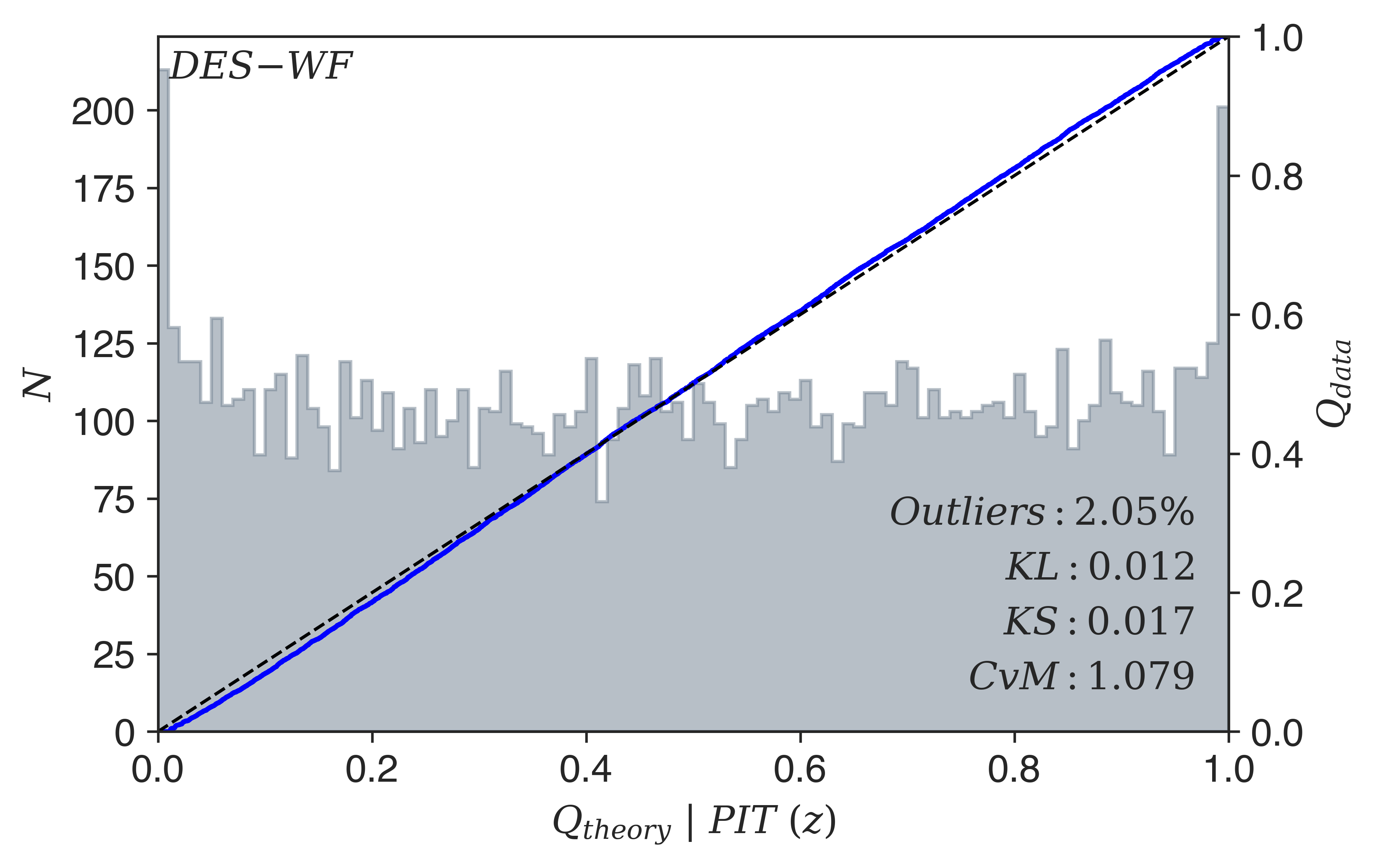}
  \end{subfigure}
  \begin{subfigure}[h!]{0.48\textwidth}
    \includegraphics[width=\textwidth]{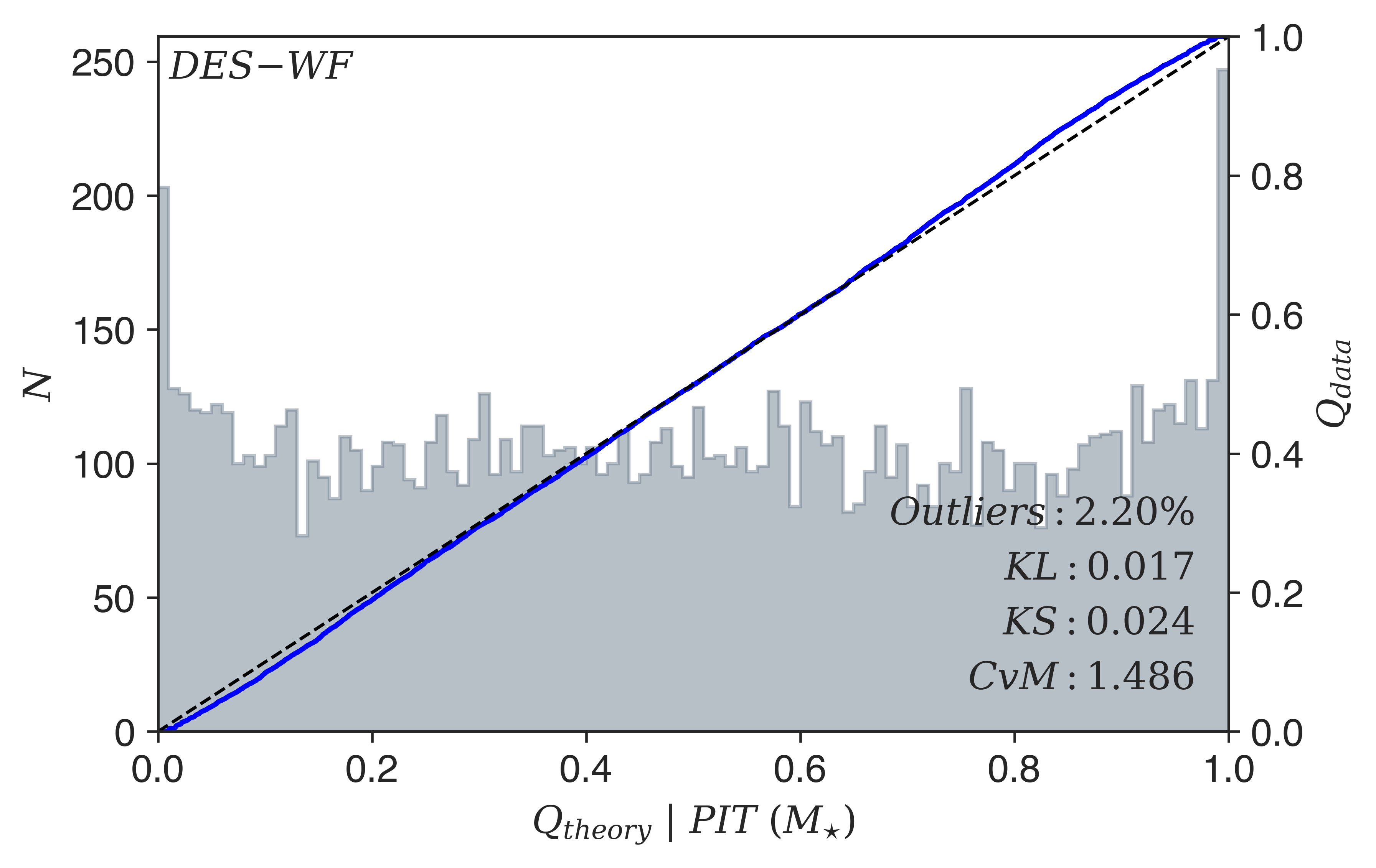}
  \end{subfigure}
  \caption{Redshift and stellar mass PIT distributions for the DES-DF and DES-WF models. These distributions are used to assess the probabilistic calibration of marginal PDFs of test galaxies produced by the models. They are overlaid with Q--Q plots to highlight deviations from uniformity. The black-dashed and solid blue lines represent the quantiles of $U(0,1)$ and PIT distributions, respectively. The percentage of catastrophic outliers along with the values of the Kullback--Leibler (KL) divergence, Kolmogorov--Smirnov (KS) test, and Cram\'er-von Mises (CvM) metrics are also stated to quantify uniformity of the PIT distributions. We define a catastrophic outlier to be any galaxy with a redshift or stellar mass completely outside the support of its marginal PDF.}
  \label{fig:pit}
\end{figure*}

\cite{gneiting_2007} introduce three modes of calibration: probabilistic, marginal, and exceedance. The first two modes are the most important, and they can be empirically assessed. As a result, we focus on these to determine if the marginal PDFs produced by our models are valid and exclude exceedance calibration in our analysis. Probabilistic calibration can be assessed using the probability integral transform (PIT; \citealt{rosenblatt}). It is the cumulative distribution function (CDF) evaluated at its true redshift or stellar mass:

\begin{equation}
    PIT \equiv \int_{-\infty}^{\Tilde{y}} f(y) dy,
\end{equation}

\noindent
where $\Tilde{y}$ is the `true' redshift or the stellar mass and $f(y)$ is the marginal PDF. If the marginal PDFs are probabilistically calibrated, then the true redshifts and stellar masses should be random draws from their respective distributions. This statement is equivalent to requiring that the CDF evaluated at the true redshift should not have a preferred value. In this case, for an ensemble of galaxies, the distribution of PIT values should follow the standard uniform distribution ($U (0, 1)$; \citealt{dawid}), i.e. one percent of galaxies should have their spec-\textit{z}s found within the first percentile of their CDFs, and so on. Deviations from uniformity can be interpreted as follows. If the marginal PDFs are overly broad, then fewer objects will have true redshifts in the tails of their PDF, instead being closer to 0.5, and the PIT distribution will be convex shaped. Conversely, if they are overly narrow, then the PIT distribution will be concave shaped. Finally, if the PIT distribution has a gradient, then this means that the marginal PDFs are biased. In the past, the PIT distribution has been utilized to determine the validity of redshift PDFs (e.g. \citealt{bordoloi}; \citealt{polsterer}; \citealt{tanaka}; \citealt{schmidt}; \citealt{desprez}).

\begin{figure*}
    \begin{subfigure}[h!]{0.48\textwidth}
    \includegraphics[width=\textwidth]{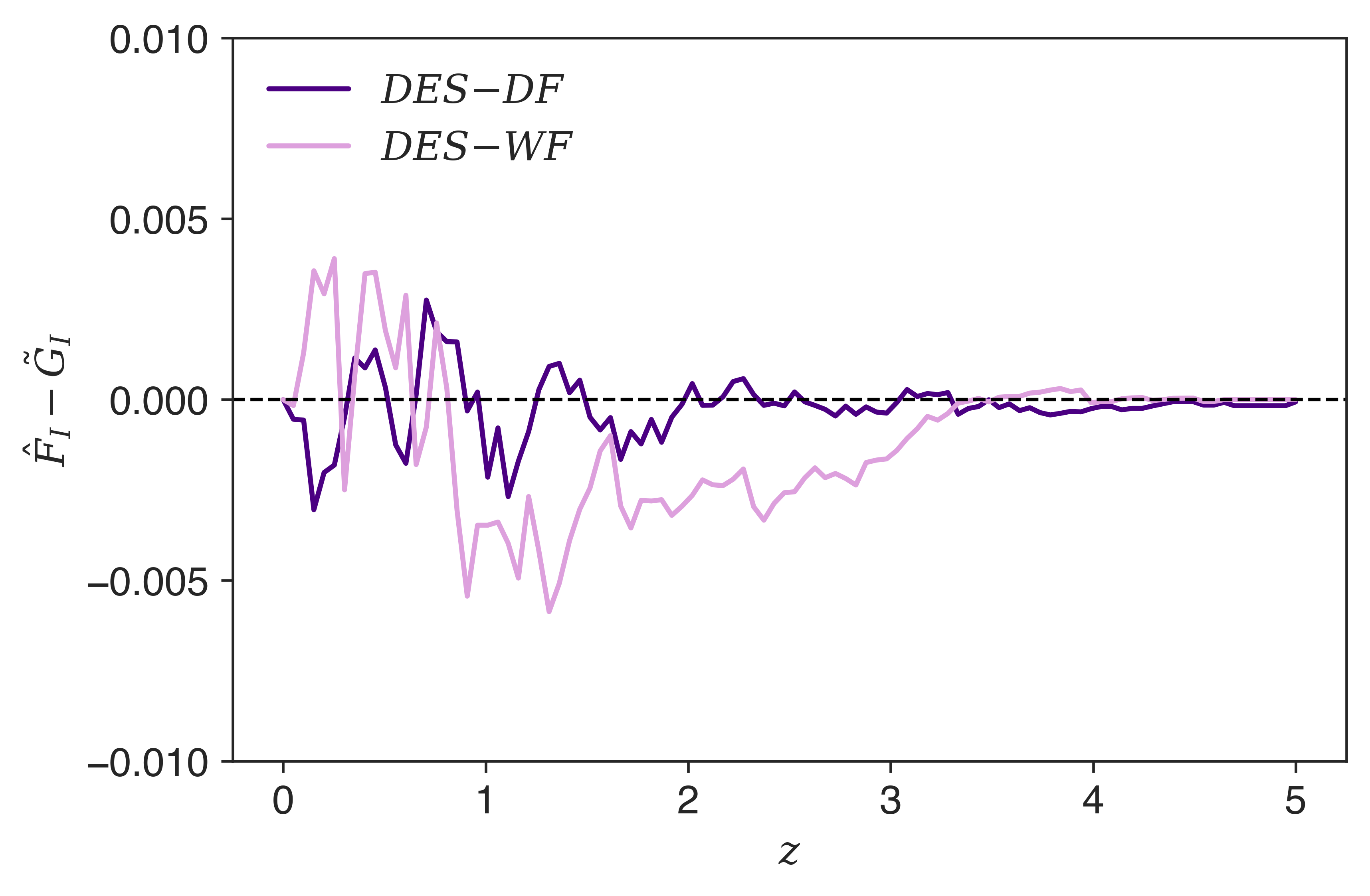}
    \end{subfigure}
    \begin{subfigure}[h!]{0.48\textwidth}
    \includegraphics[width=\textwidth]{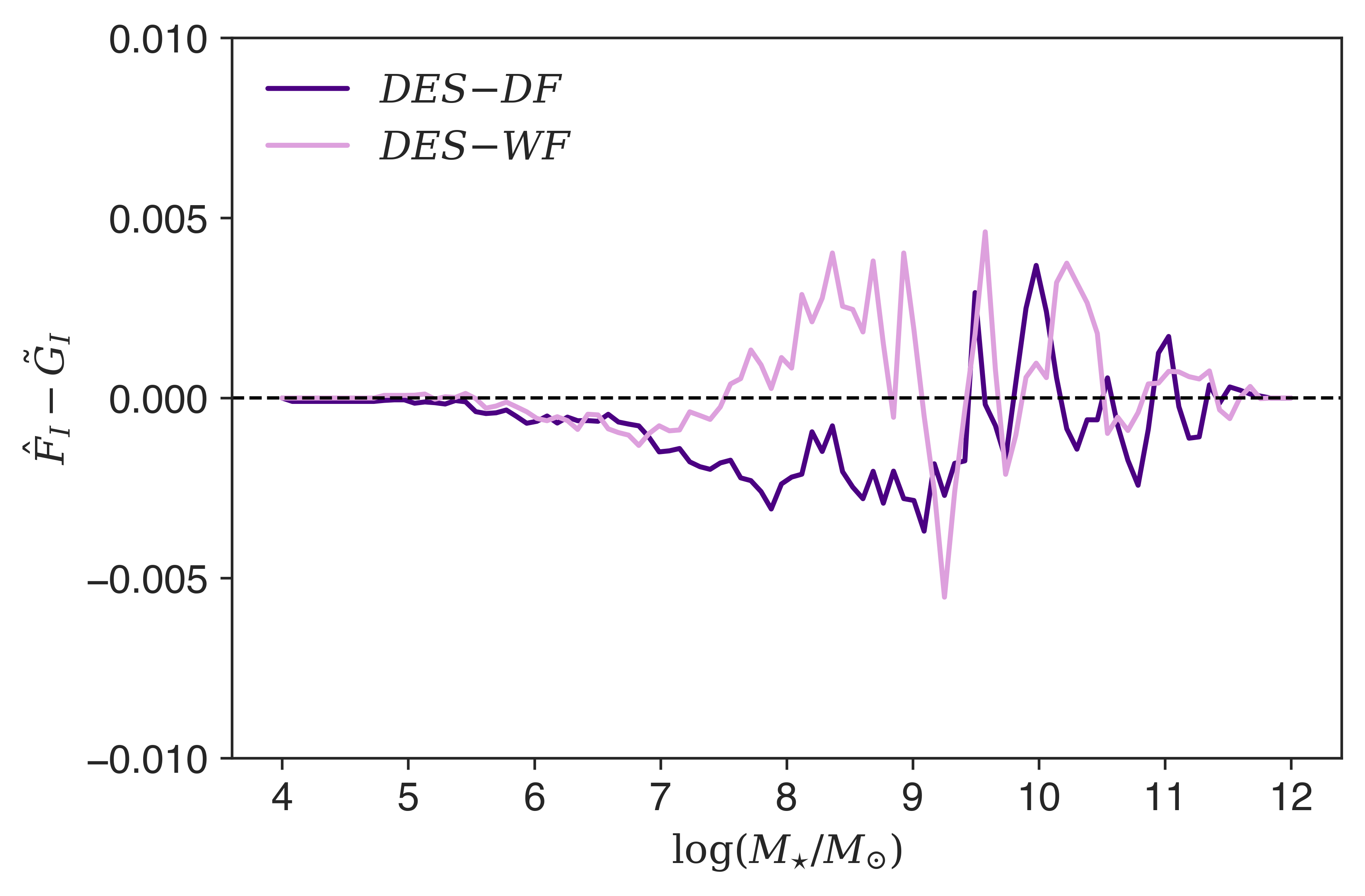}
    \end{subfigure}
\caption{The difference between the average predictive CDF ($\hat{F}_{I}$) and the true empirical CDF ($\Tilde{G}_{I}$) of redshift and stellar mass plotted at different intervals in their respective ranges. These diagnostic plots are used to assess the marginal calibration of marginal PDFs of test galaxies produced by the DES-DF and DES-WF models.}
\label{fig:marginal_calibration}
\end{figure*}

The uniformity of the PIT distribution is a necessary condition for marginal PDFs to be valid. However, \cite{hamill} has shown that uniformity can also arise from biased distributions. Therefore, probabilistic calibration may not be sufficient in some cases, and marginal calibration may be required to reach a concrete conclusion. Marginal calibration is associated with the equality of the predicted and true distributions of redshift and stellar mass. Specifically, the average predictive CDF ($\hat{F}_{I}$) is compared to the true empirical CDF ($\Tilde{G}_{I}$).

\begin{equation}
    \hat{F}_{I}(y) = \frac{1}{n} \sum_{i=1}^{n} F_{i}(y),
\end{equation}

\begin{equation}
    \Tilde{G}_{I}(y) = \frac{1}{n} \sum_{i=1}^{n} \mathbbm{1}\{\Tilde{y}_{i} \leq y\},
\end{equation}

\noindent
where $n$ is the number of test galaxies, $F_{i}$ is the predicted CDF, $\Tilde{y}_{i}$ is the true redshift or the stellar mass of a galaxy, and $\mathbbm{1}$ is the indicator function, defined as

\begin{equation}
    \mathbbm{1}\{\Tilde{y}_{i} \leq y\} =\begin{cases}
    1 & \text{if True}\\
    0 & \text{if False}
  \end{cases}.
\end{equation}

\noindent
If the PDFs are marginally calibrated, then the average predictive CDF should equal the true empirical CDF. To assess probabilistic calibration, we check the uniformity of the PIT distributions visually and use quantile--quantile (Q--Q) plots to highlight deviations. In a Q--Q plot, the quantiles of one distribution are plotted against the quantiles of another distribution. In our case, these are the PIT and $U (0, 1)$. If the two distributions are identical, then the quantiles match and lie along the diagonal. Furthermore, we use several metrics to quantitatively determine the uniformity of the PIT distributions \citep{schmidt} such as the Kullback--Leibler (KL; \citealt{kl_divergence}) divergence, Kolmogorov--Smirnov (KS; \citealt{ks_test}) test and Cram\'er-von Mises (CvM; \citealt{cramer}) test. All of these metrics measure the similarity between two distributions in different ways. The KL divergence is defined by the following integral:

\begin{equation}
    KL \equiv \int_{-\infty}^{\infty} p(x) \log(\frac{p(x)}{q(x)}) dx,
\end{equation}

\noindent
where $p(x)$ and $q(x)$ are the reference ($U (0, 1)$) and target (PIT) PDFs, respectively. The KS test is a non-parametric test and is the maximum distance between the empirical distribution function ($F_{n}(x)$) and the CDF ($F(x)$) of the reference distribution:

\begin{equation}
    KS \equiv sup_{x}|F_{n}(x) - F(x)|,
\end{equation}

\noindent
where $sup_{x}$ is the supremum of the set of distances. The CvM is an alternative to KS test and is more sensitive to the edges of a distribution:

\begin{equation}
    CvM \equiv \int_{-\infty}^{\infty}
    (F_{n}(x) - F(x))^{2} dF(x).
\end{equation}

\noindent
A value of zero for the different metrics indicates that there is a perfect match between the two distributions. 

Fig. \ref{fig:pit} shows the redshift and stellar mass PIT distributions and Q--Q plots for the models. The black-dashed line represents the quantiles of $U(0,1)$, and the quantiles of the PIT distributions are shown using the solid blue curves. The values of the metrics, along with the percentage of catastrophic outliers, are also indicated. We define a catastrophic outlier to be any galaxy for which the true value of redshift or stellar mass is completely outside the support of its marginal PDF.

Visually, the PIT distributions of DES-DF and DES-WF appear to be uniform, and this is reinforced by the quantiles of the PIT distributions lying close to the diagonal in the Q--Q plots, if not on it. Consequently, at first glance, both models seem to be performing equally well. However, on closer inspection, subtle differences can be observed in the PIT distributions. The PIT distributions of DES-DF are more uniform compared to those of DES-WF, and the main difference arises at the edges. Specifically, the PIT distributions of DES-WF are slightly concave shaped as indicated by the minor deviations in the Q--Q plots at the extremes and quantitatively confirmed by the significantly larger CvM criterion values. Hence, the marginal PDFs produced by DES-WF are somewhat overly narrow or underdispersed. Taking into account the degraded photometry, DES-WF is still performing admirably with only small increases in the number of catastrophic outliers compared to DES-DF. Overall, both models are producing probabilistically calibrated marginal PDFs and performing at an unprecedented level.

\begin{figure*}
  \begin{subfigure}[h!]{0.41\textwidth}
    \includegraphics[width=\textwidth]{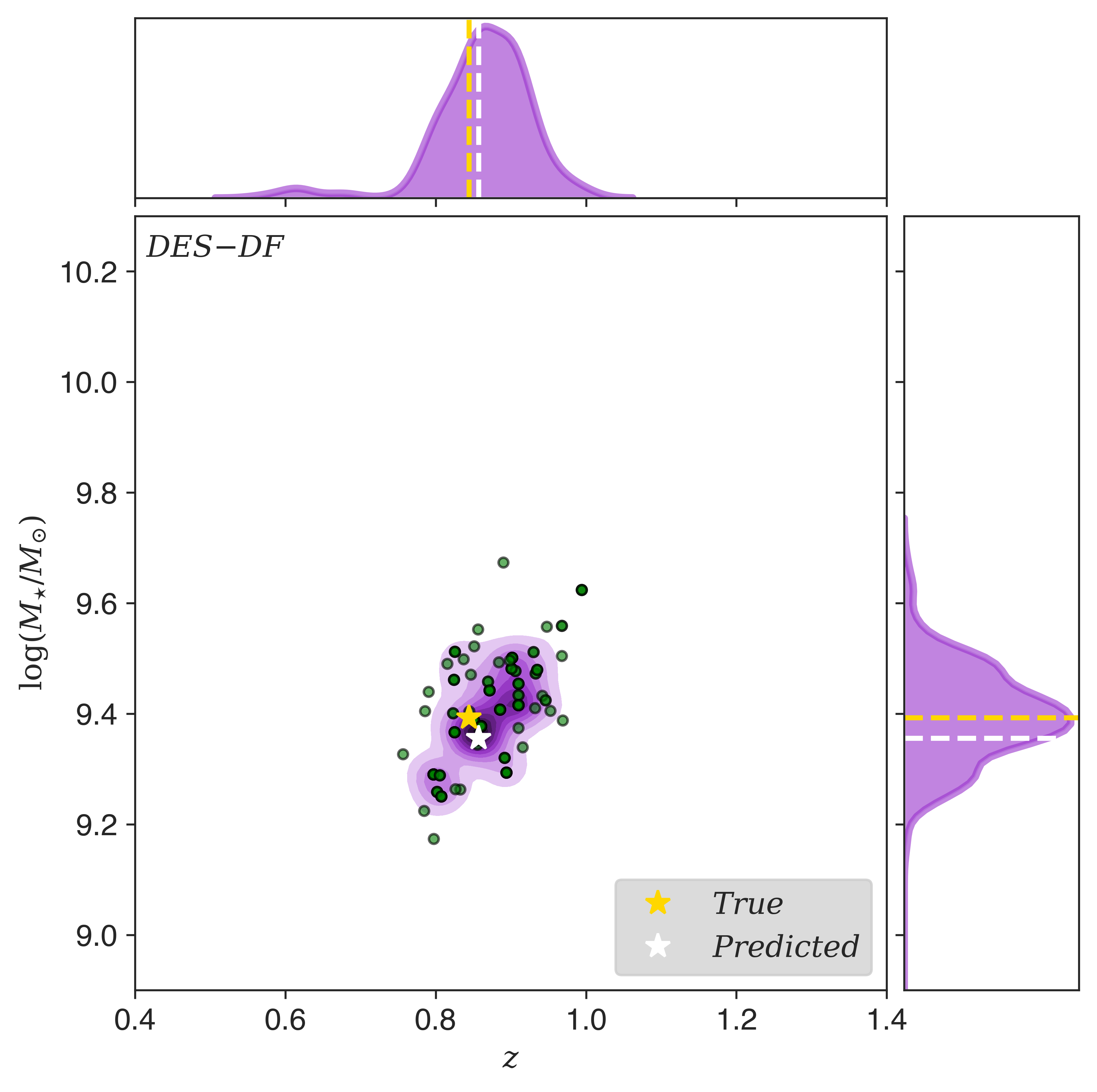}
  \end{subfigure}
  \begin{subfigure}[h!]{0.41\textwidth}
    \includegraphics[width=\textwidth]{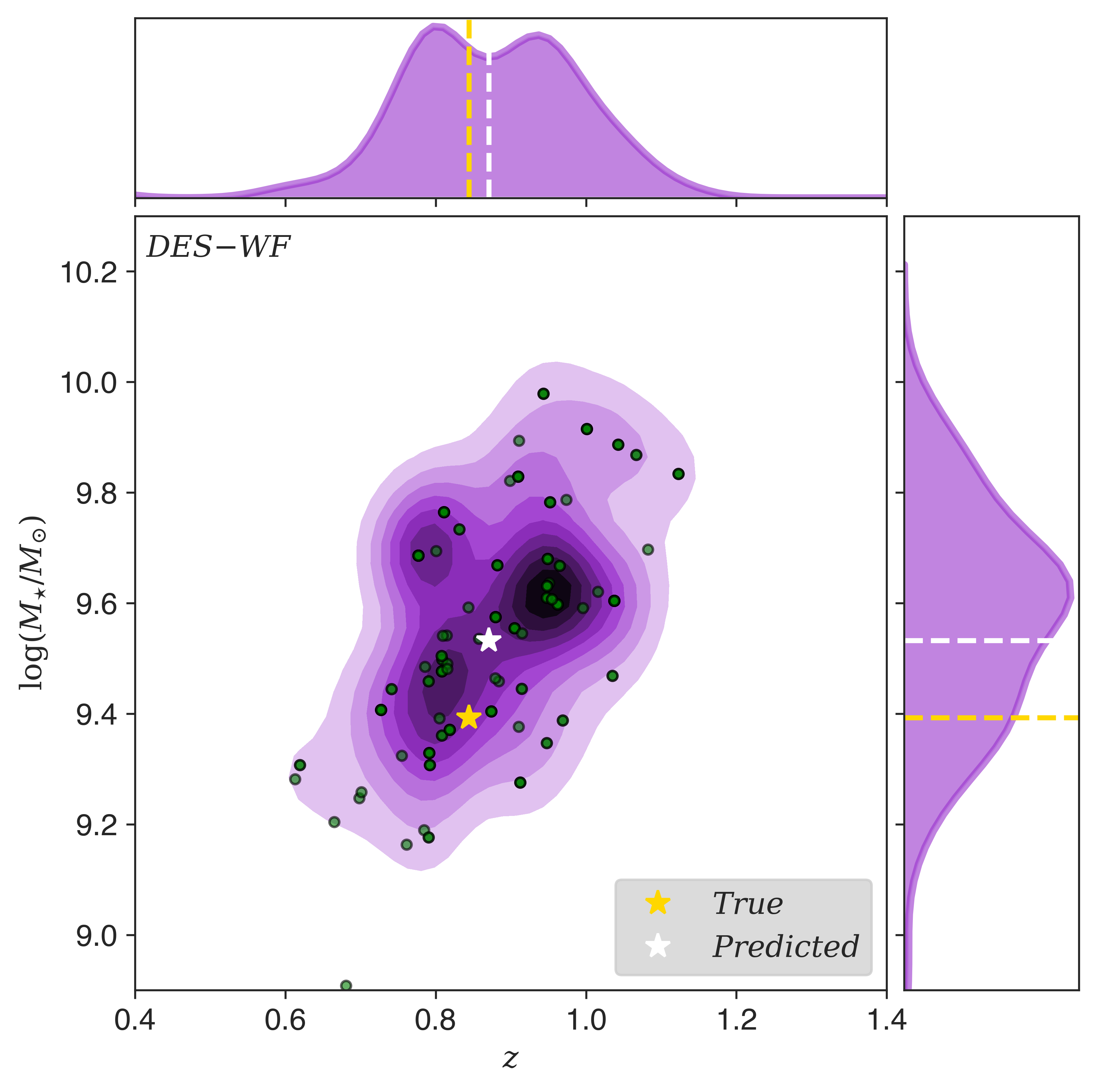}
  \end{subfigure}
    \begin{subfigure}[h!]{0.41\textwidth}
    \includegraphics[width=\textwidth]{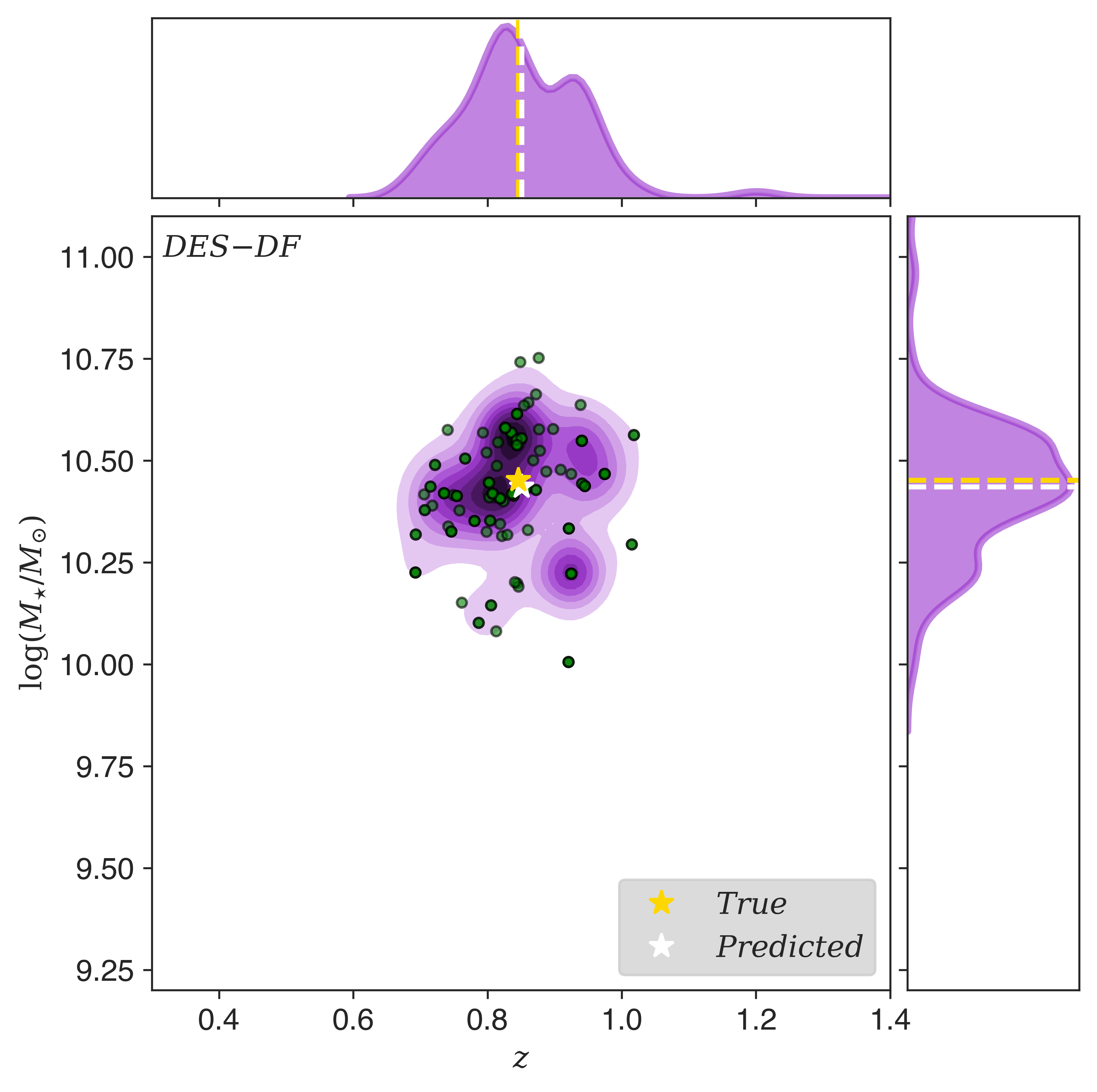}
  \end{subfigure}
  \begin{subfigure}[h!]{0.41\textwidth}
    \includegraphics[width=\textwidth]{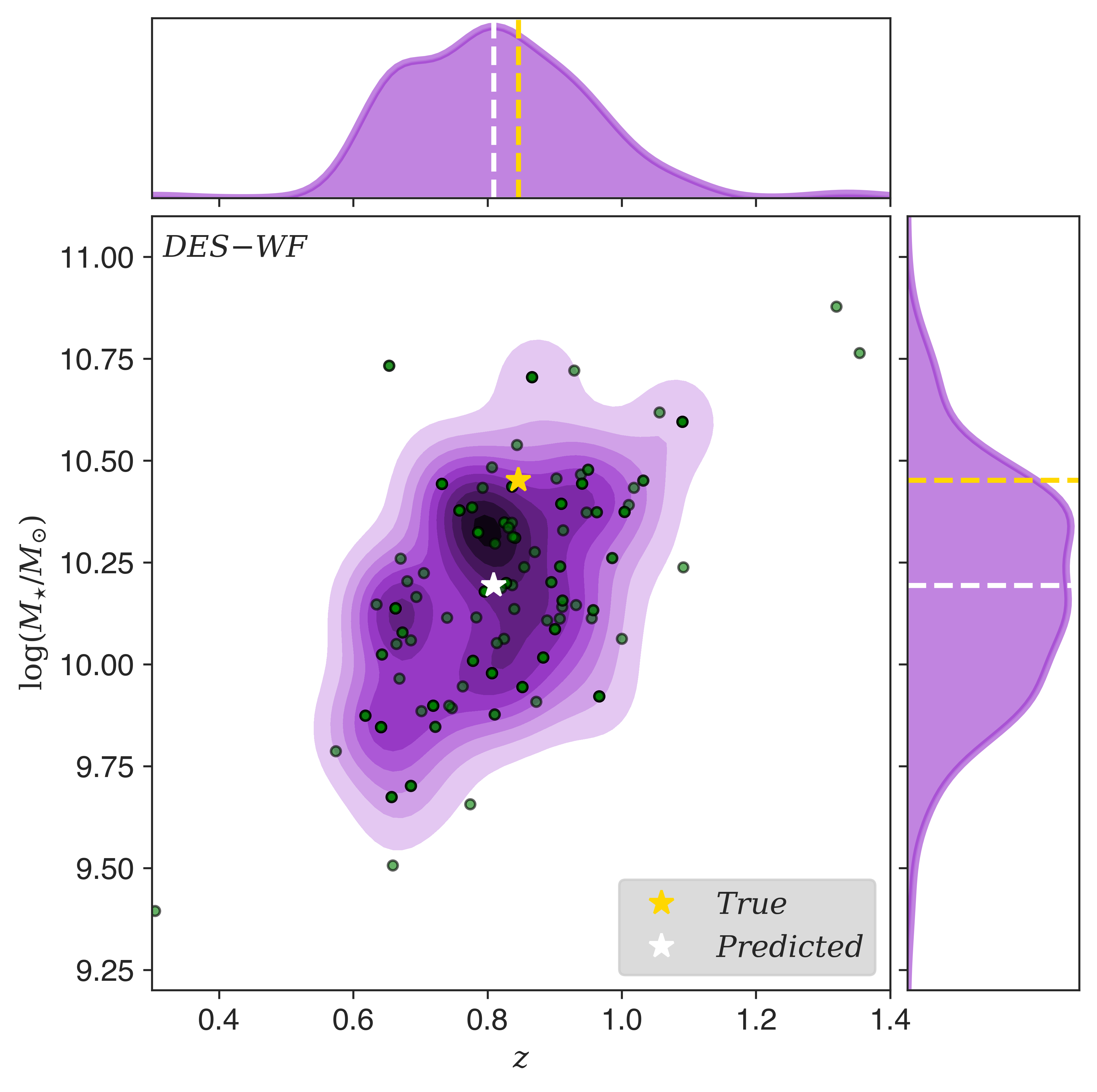}
  \end{subfigure}
    \begin{subfigure}[h!]{0.41\textwidth}
    \includegraphics[width=\textwidth]{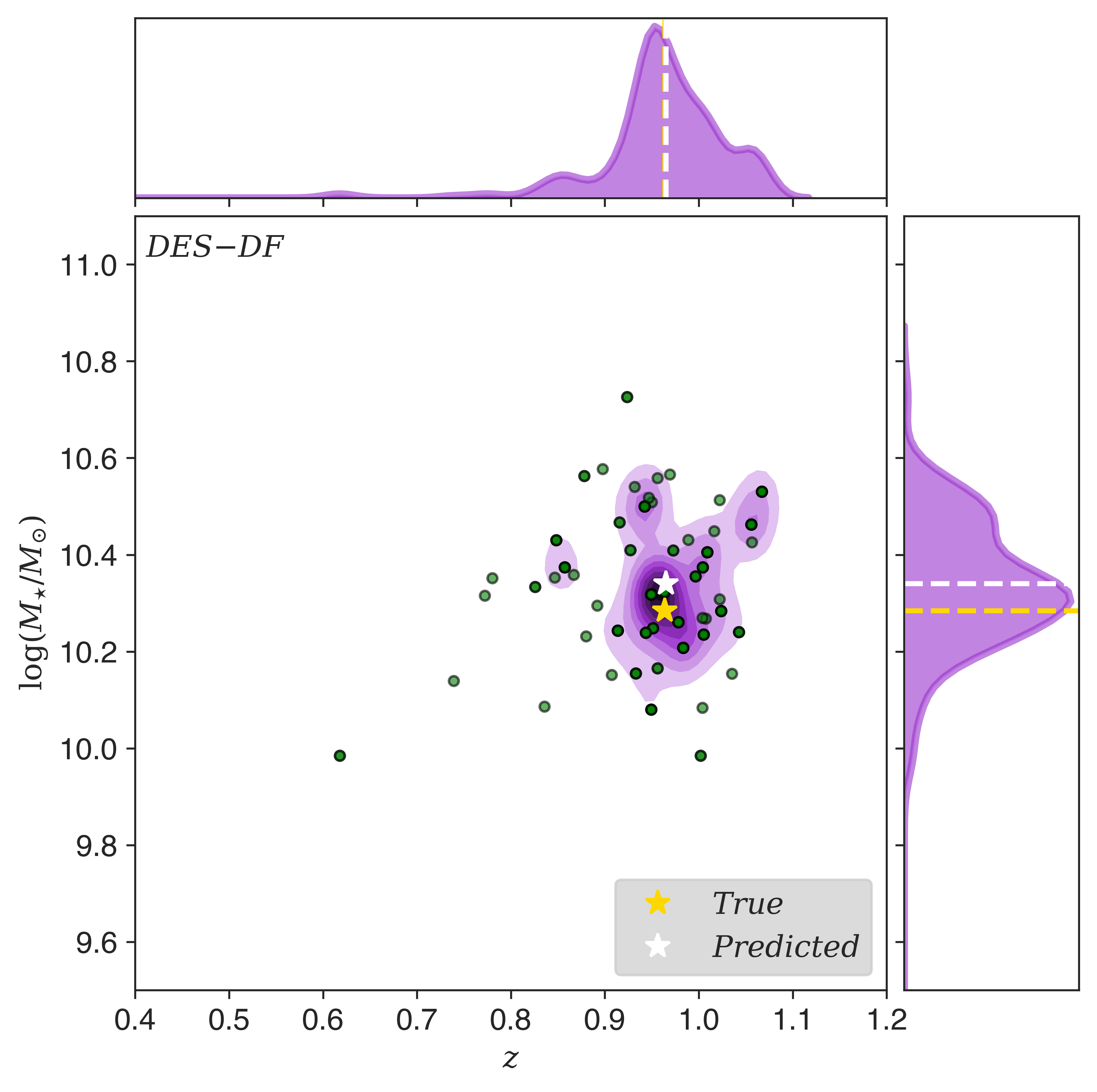}
  \end{subfigure}
  \begin{subfigure}[h!]{0.41\textwidth}
    \includegraphics[width=\textwidth]{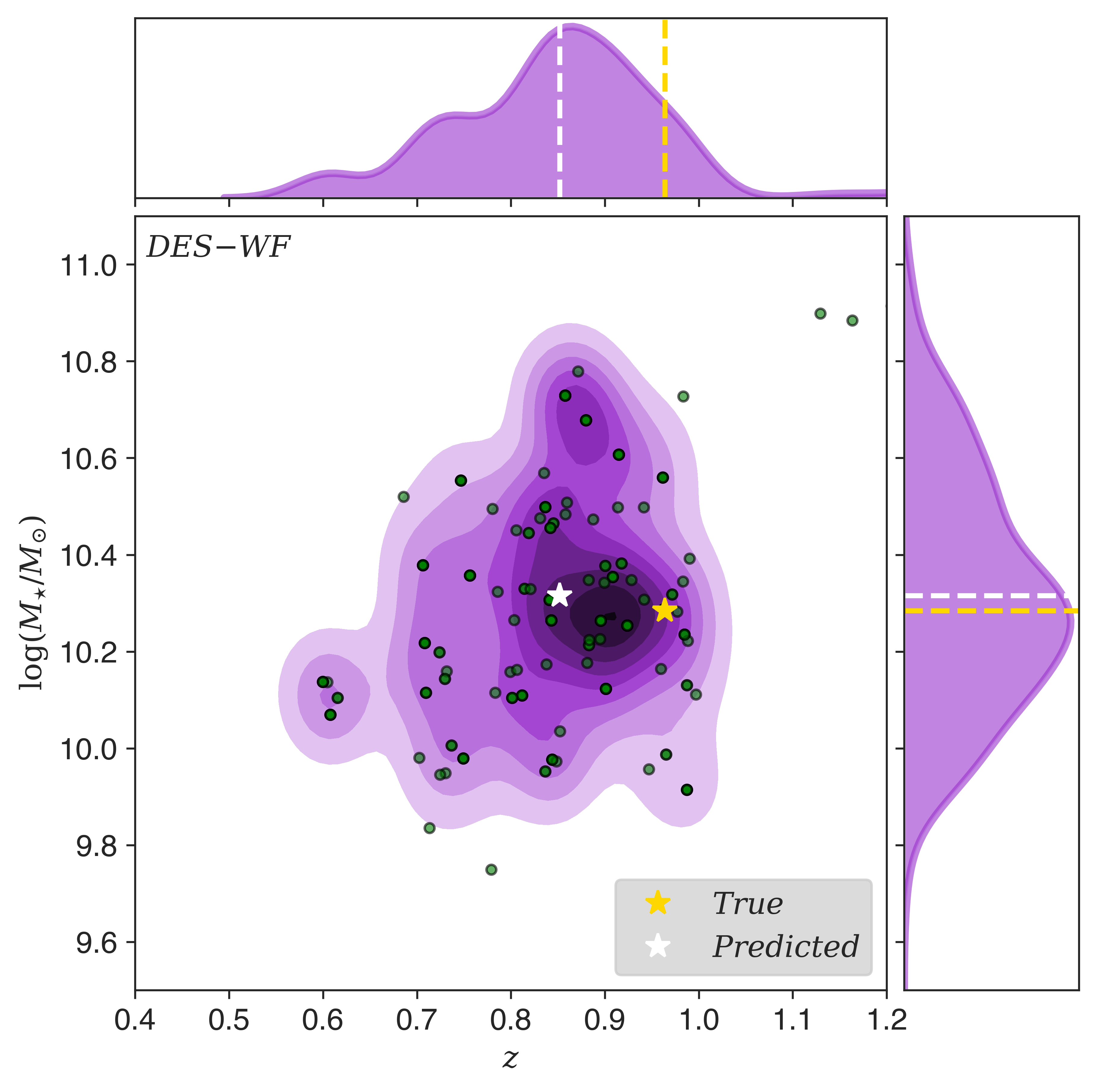}
  \end{subfigure}
  \caption{Examples of joint redshift--stellar mass PDFs produced by the DES-DF and DES-WF models of the same test galaxies (in rows). The gold and white stars alongside the dashed lines represent the `true' and predicted redshifts and stellar masses of the galaxies respectively. The predicted redshifts and stellar masses are computed by averaging the predictions from all the decision trees in the individual RFs. The green circles indicate the values of redshift and stellar mass in the leaf nodes that are representative of the test galaxies.}
  \label{fig:joint_pdfs}
\end{figure*}

\begin{figure*}
  \begin{subfigure}[h!]{0.48\textwidth}
    \includegraphics[width=\textwidth]{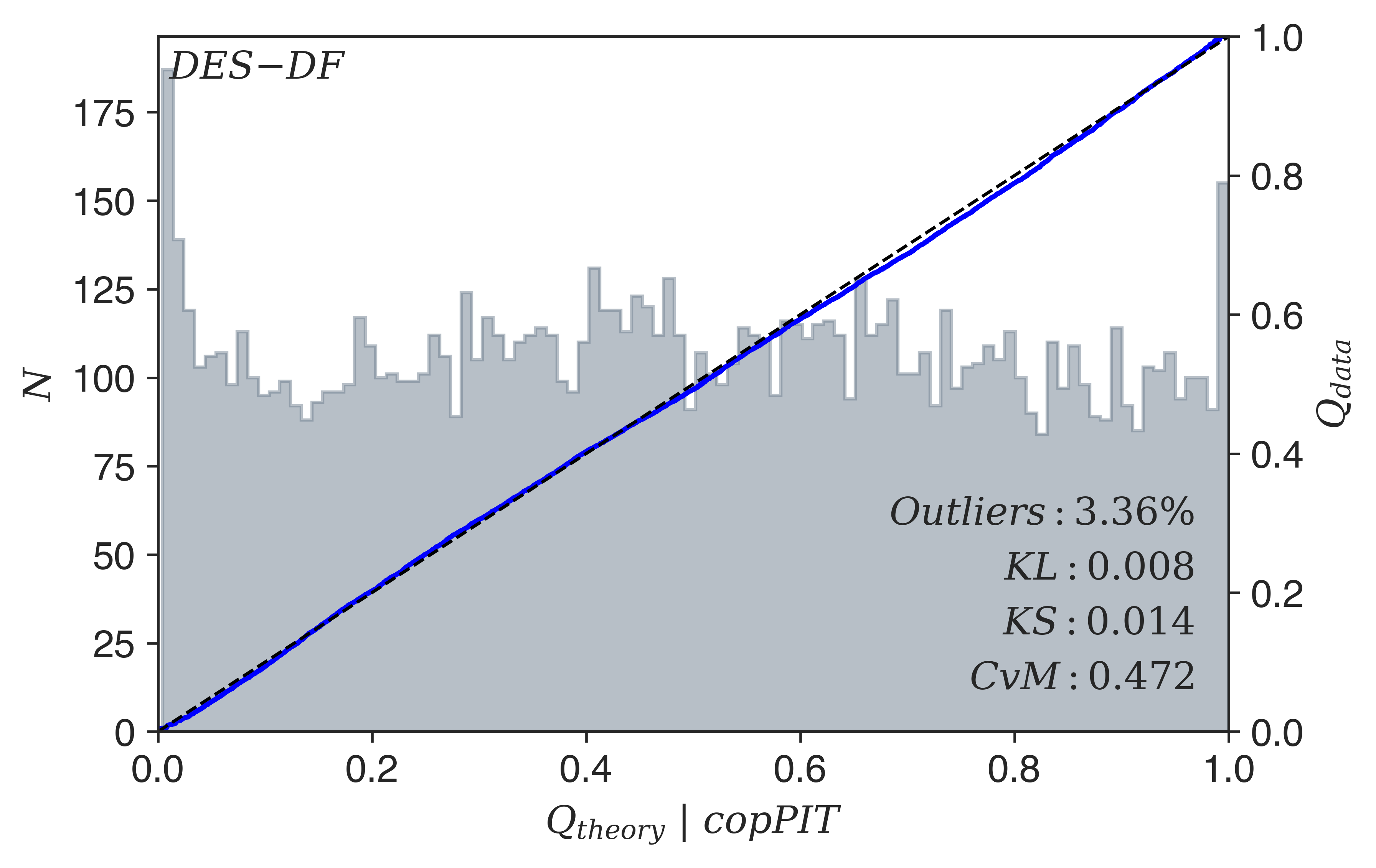}
  \end{subfigure}
  \begin{subfigure}[h!]{0.48\textwidth}
    \includegraphics[width=\textwidth]{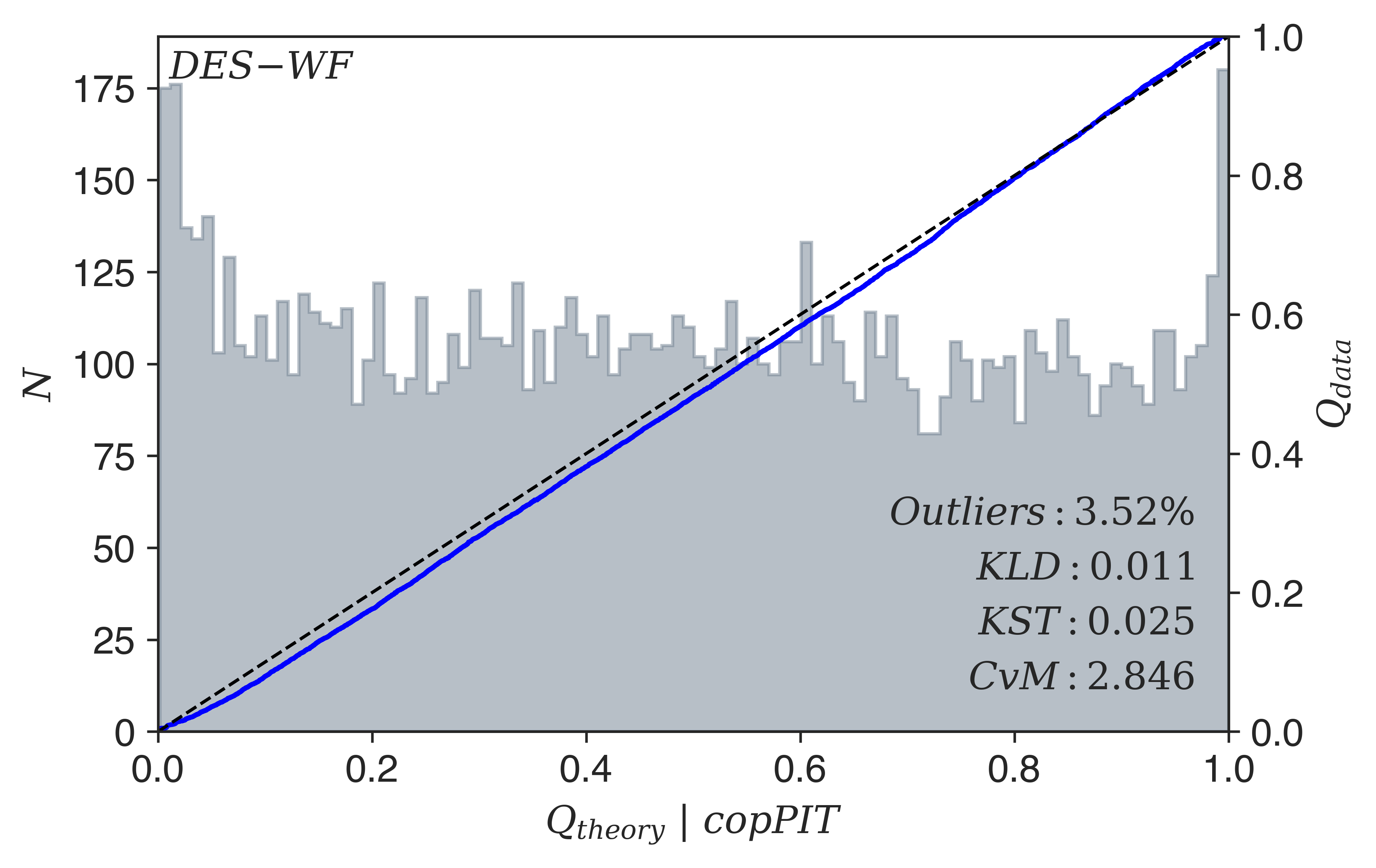}
  \end{subfigure}
  \caption{copPIT distributions for the DES-DF and DES-WF models. They are overlaid with Q--Q plots to aid in visually assessing the probabilistic copula calibration of joint redshift--stellar mass PDFs of test galaxies. The black-dashed and solid blue lines represent the quantiles of $U(0,1)$ and copPIT distributions, respectively. The percentage of catastrophic outliers along with the values of the Kullback--Leibler (KL) divergence, Kolmogorov--Smirnov (KS) test, and Cram\'er-von Mises (CvM) metrics is also stated to quantify uniformity of the copPIT distributions. We define a catastrophic outlier to be any galaxy that is completely outside the support of its marginal PDFs. Probabilistic copula calibration is the multivariate analogue of probabilistic calibration.}
  \label{fig:coppit}
\end{figure*}

To assess marginal calibration, we plot the difference between the average predictive and true empirical CDFs of redshift and stellar mass at regular intervals in their respective ranges. If the PDFs are marginally calibrated, then only minor fluctuations about the zero line are expected. Fig. \ref{fig:marginal_calibration} shows the redshift and stellar mass marginal calibration for the models. There are negligible fluctuations about the zero line, with maximum deviations of $\sim 0.005$. Therefore, both models are producing marginally calibrated redshift and stellar mass PDFs, with DES-DF performing marginally better with a smaller average deviation compared to DES-WF. To summarize, the marginal PDFs are both probabilistically and marginally calibrated, thus giving us confidence that they are valid. Finally, in the next section we analyse and perform validation checks of the joint redshift--stellar mass posterior distributions.

\subsection{Joint probability distributions}
\label{subsection:joint_pdfs}
In general, a joint PDF encompasses more information than its marginals. Therefore, we extract joint redshift--stellar mass PDFs of test galaxies from DES-DF and DES-WF. We build the distributions by combining the aggregated values of redshift and stellar mass in the leaf nodes across all the decision trees. Fig. \ref{fig:joint_pdfs} shows some examples of the joint PDFs of the same test galaxies produced by the models. The gold and white stars alongside the dashed lines indicate the `true' and predicted redshifts and stellar masses, respectively. We remind the reader that the predicted redshifts and stellar masses are computed by averaging the predictions from all the decision trees in a RF. Visually, the joint PDFs of the same test galaxy look alike and occupy similar regions of the redshift--stellar mass space. However, the joint PDFs produced by DES-WF are more spread out compared to the ones produced by DES-DF, or in other words, the probability is more dispersed. This is a reflection of the degraded photometry in the WF data set. Overall, we do not expect the joint PDFs of the same galaxy to resemble each other perfectly as both models have been trained using different data sets.

\subsubsection{Joint PDFs validation}
\label{subsection:joint_pdfs_validation}
It is more challenging to validate joint PDFs compared to marginal PDFs as the relatively straightforward methods adopted to validate the latter are no longer applicable. As a result, we use the multivariate extensions of probabilistic and marginal calibration developed by \cite{ziegel} to validate joint PDFs in our case. These are probabilistic copula calibration and Kendall calibration, respectively. Probabilistic copula calibration can be empirically assessed by using the copula probability integral transform (copPIT):

\begin{figure}
    \centering
    \includegraphics[width=\columnwidth]{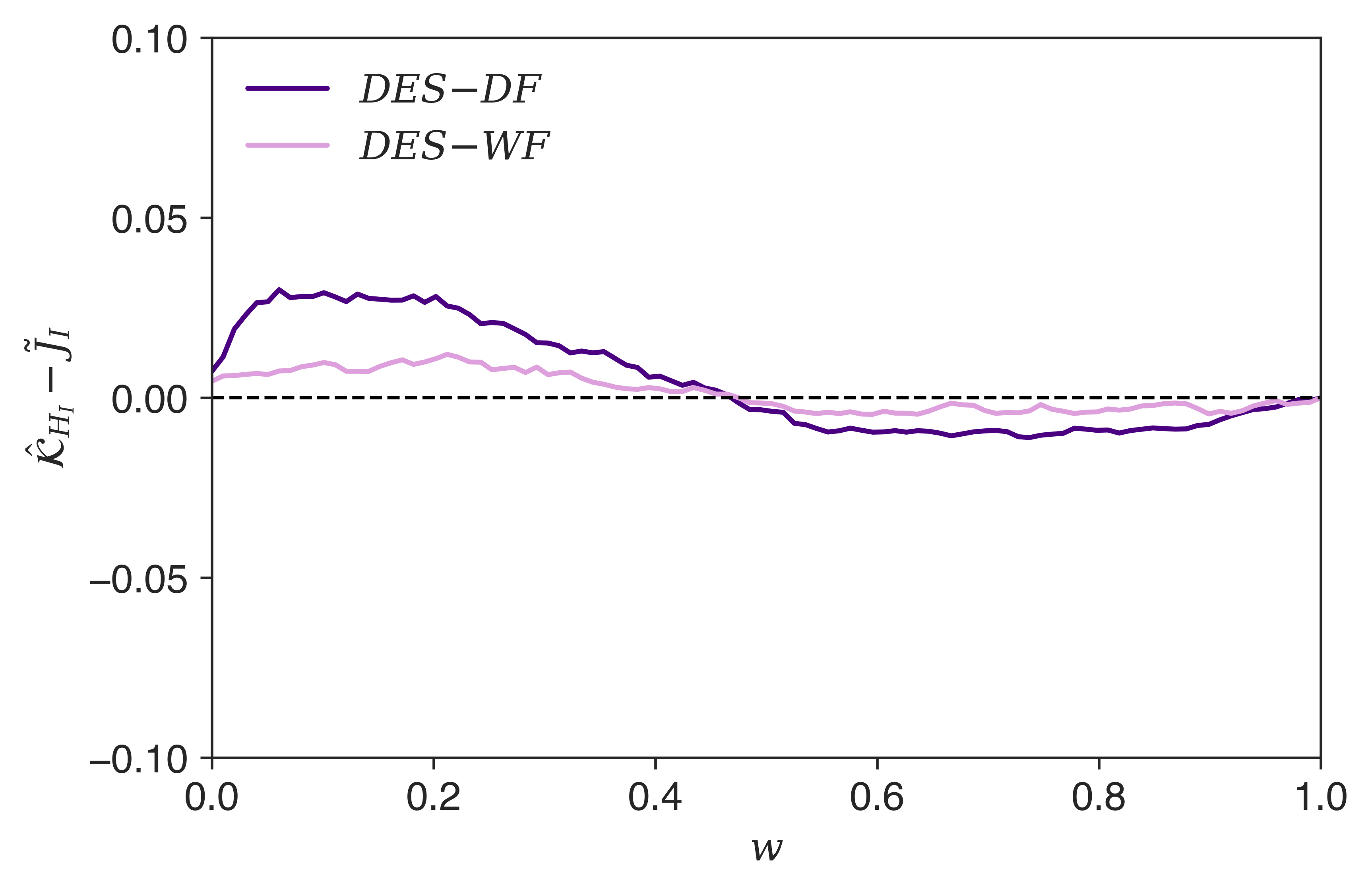}
\caption{The difference between the `average Kendall distribution function' ($\mathcal{\hat{K}}_{H_{I}}$) and the empirical CDF of the predicted joint CDFs evaluated at the `true' redshifts and stellar masses ($\Tilde{J}_{I}$), plotted at regular intervals in the probability space $w \in [0, 1]$. This diagnostic plot is used to assess the Kendall calibration of the joint PDFs produced by the DES-DF and DES-WF models. Kendall calibration is the multivariate analogue of marginal calibration.}
\label{fig:kendall_calibration}
\end{figure}

\begin{equation}
    copPIT \equiv \mathcal{K}_{H}(H(\Tilde{y})),
\end{equation}

\noindent
where $H(\Tilde{y})$ is the joint CDF evaluated at the true redshift and stellar mass, and $\mathcal{K}_{H}$ is the Kendall distribution function, defined as

\begin{equation}
    \mathcal{K}_{H}(w) = P (H(y) \leq w),
\end{equation}

\noindent
where $H(y)$ is the predicted joint CDF and $w \in [0, 1]$. Simply put, the Kendall distribution function is the CDF of $H(y)$. For marginal PDFs, it corresponds to the standard uniform distribution and the copPIT coincides with the PIT. To assess Kendall calibration, we compare what we refer to as the `average Kendall distribution function' ($\mathcal{\hat{K}}_{H_{I}}$) to the empirical CDF of the predicted joint CDFs evaluated at the `true' redshifts and stellar masses ($\Tilde{J}_{I}$):

\begin{equation}
    \mathcal{\hat{K}}_{H_{I}} (w) = \frac{1}{n} \sum_{i=1}^{n} \mathcal{K}_{H_{i}} (w),
\end{equation}

\begin{equation}
   \Tilde{J}_{I}(w) = \frac{1}{n} \sum_{i=1}^{n} \mathbbm{1} \{H_{i}(\Tilde{y}_{i}) \leq w\}.
\end{equation}

Probabilistic copula calibration and Kendall calibration can be interpreted in the same manner as their univariate counterparts. As such, probabilistic copula calibration ascertains if the `true' redshifts and stellar masses of galaxies are random draws from their corresponding joint PDFs, as they should be.  If this is the case, then for an ensemble, the copPIT distribution is uniform, and the joint PDFs are probabilistically copula calibrated. On the other hand, Kendall calibration probes how well the dependence structure between redshift and stellar mass is predicted on average, and can be understood as marginal calibration of the Kendall distribution. If $\mathcal{\hat{K}}_{H_{I}}$ is comparable to $\Tilde{J}_{I}$, then the joint PDFs are Kendall calibrated. Once again, if both modes of calibration are satisfied, then we can claim with some conviction that the joint PDFs are valid overall. Furthermore, we would like to point out that while we use probabilistic copula calibration and Kendall calibration to validate our joint redshift--stellar mass PDFs, they can be applied to validate higher dimensional PDFs also. 

Fig. \ref{fig:coppit} shows the copPIT distributions for the DES-DF and DES-WF models. The distributions are uniform with minor deviations that are more prominent for DES-WF. Overall, both models are performing well with no substantial differentiation and producing joint PDFs that are probabilistically copula calibrated. Furthermore, in comparison to the PIT distributions in Fig. \ref{fig:pit}, the copPIT distributions of DES-WF are somewhat less uniform as primarily reflected by the large CvM value. Hence, the marginal PDFs produced by the model are better probabilistically calibrated than the joint PDFs.

Fig. \ref{fig:kendall_calibration} shows the difference between $\mathcal{\hat{K}}_{H_{I}}$ and $\Tilde{J}_{I}$ at regular intervals in the probability space $w$. For DES-WF, the fluctuations about the zero line are smaller compared to those for DES-DF, thus indicating that the joint PDFs produced by the former are better Kendall calibrated. We believe that DES-WF is better capturing the redshift--stellar mass dependence structure as it is trained using the WF data set that contains multiple scattered copies of the same DF galaxies, resulting in better incorporation of photometric errors present in the data into the model. Collectively, the joint PDFs are less marginal/Kendall calibrated compared to the marginal PDFs as the deviations are larger in magnitude. However, we hypothesize that the deviations in the Kendall calibration are not significant given the complex nature of joint PDFs, and to prove this, we compare our results to those achieved by the template-fitting code \texttt{BAGPIPES} in the next section.

\begin{table}
    \centering
    \begin{tabular}{cc}
    \hline
    Instrument/Telescope (Survey) & Band  \\
    \hline
    UltraVista & Y, J, H, Ks \vspace{0.2cm} \\ 
    CFHT & u  \vspace{0.2cm} \\ 
    Subaru & B, V, r, i+, z++, \\
    & IA427, IA464, IA484, IA505 \\
    & IA527, IA574, IA624, IA479 \\
    & IA709, IA738, IA767, IA827 \\ \hline
    \end{tabular}
    \caption{List of 22 COSMOS bands used to build a `truth' catalogue to validate the marginal and the joint PDFs of redshift and stellar mass produced by \texttt{BAGPIPES} using the four-band ($V$, $r$, $i+$, and $z++$) Subaru  photometry.}
    \label{tab:cosmos_bands}
\end{table}

\begin{table*}
    \centering
    \begin{tabular}{ccccccc}
    \hline
    Free parameter & Prior & Limits & & & Fixed parameter & Value \\
    \hline
    $A_{V}$ & Uniform & [0, 4] & & &  $\log_{10} (U)$ & -3 \\
    $\log_{10} (M_{\star}/M_{\odot})$ & Uniform & [4, 13] & & &
    $a_{BC}$ & 0.01 Gyr \\
    $z$ & Uniform & [0, 10] & & & $\epsilon$ & 3 \\
    $\tau$ & Uniform & [0.3, 10] & & & SPS models & \cite{bc03}  \\
    $Z$ & Uniform & [0, 2.5] & & & IMF & \cite{kroupa} \\ \hline
    \end{tabular}
    \caption{Fixed and fitted parameters with their associated priors for the delayed exponentially declining ($\tau^{-2} t e^{-t/\tau}$) star formation history (SFH) model used in the \texttt{BAGPIPES} runs. The model is not readily available in \texttt{BAGPIPES}, so we lightly modify the code to meet our requirements. We adopt the \protect\cite{calzetti} attenuation curve, stellar population synthesis (SPS) models of \protect\cite{bruzual} and a \protect\cite{kroupa} initial mass function (IMF). $A_{V}$ is the attenuation in the V-band, $\tau$ is the star formation time-scale, $Z$ is the metallicity, $U$ is the ionization parameter, $a_{BC}$ is the lifetime of H II regions and $\epsilon$ is a constant that controls the extra attenuation towards them.}
    \label{tab:priors}
\end{table*}

\begin{figure*}
  \begin{subfigure}[h!]{0.46\textwidth}
    \includegraphics[width=\textwidth]{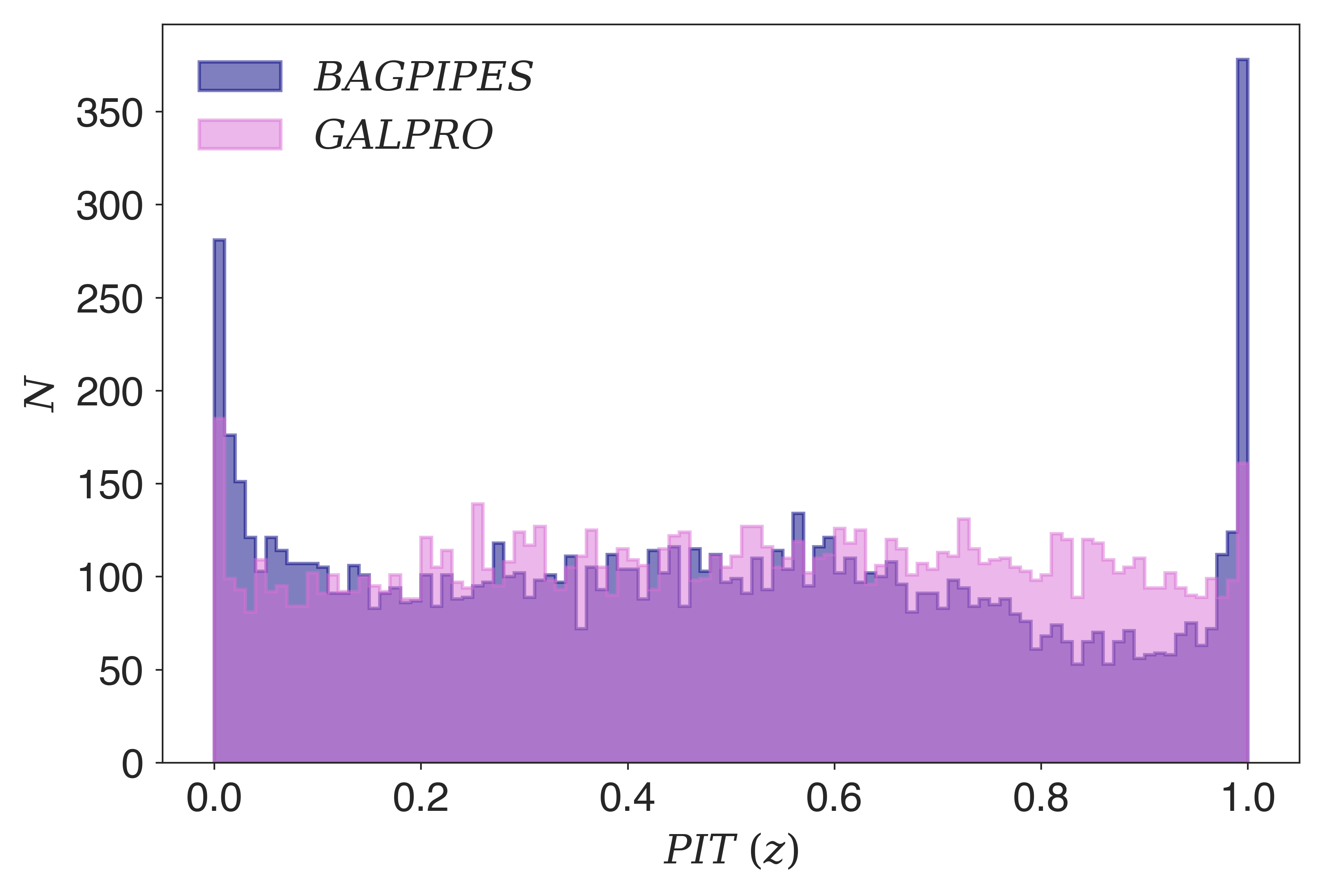}
  \end{subfigure}
 \begin{subfigure}[h!]{0.46\textwidth}
    \includegraphics[width=\textwidth]{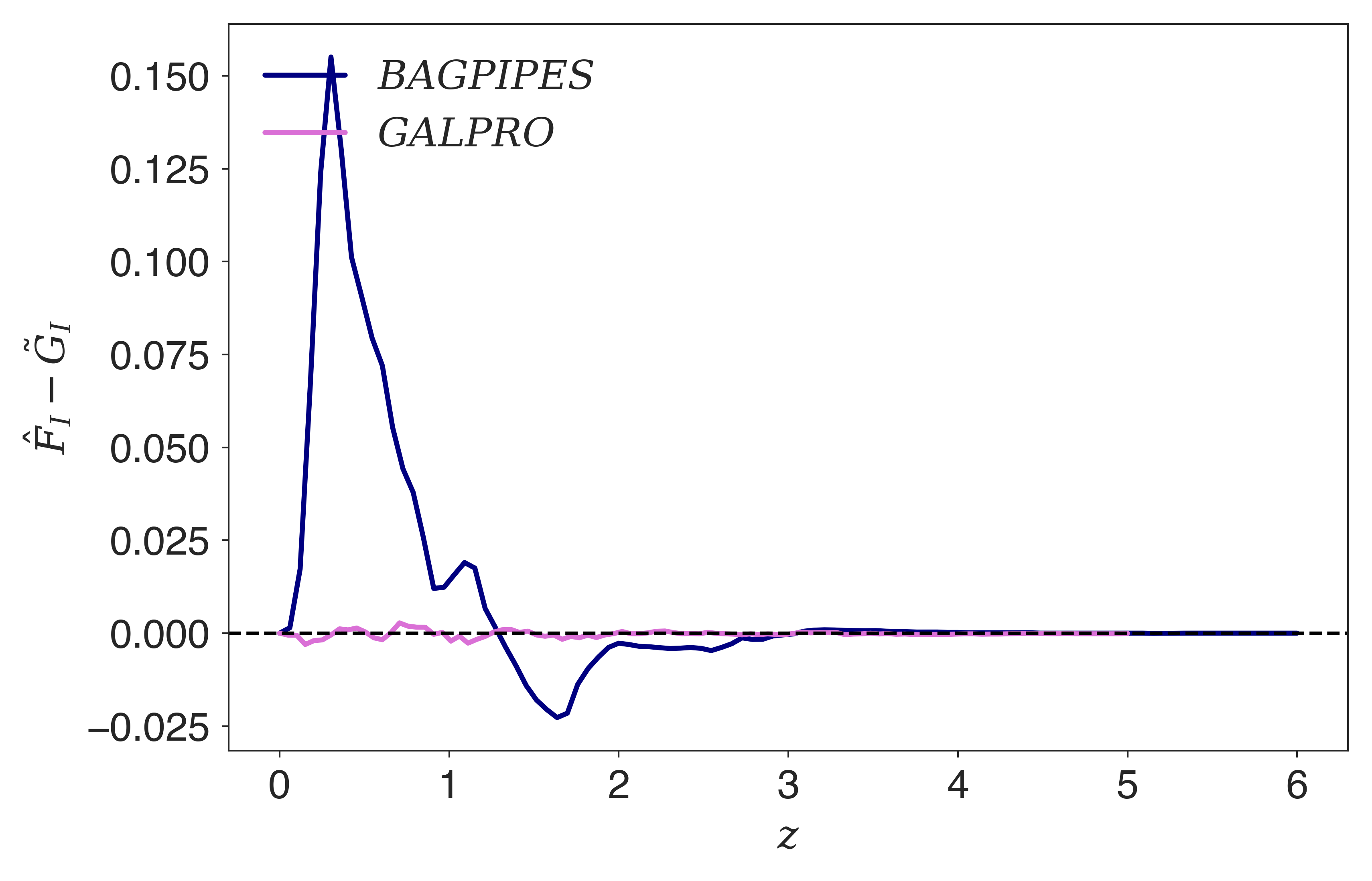}
  \end{subfigure}
  \begin{subfigure}[h!]{0.46\textwidth}
    \includegraphics[width=\textwidth]{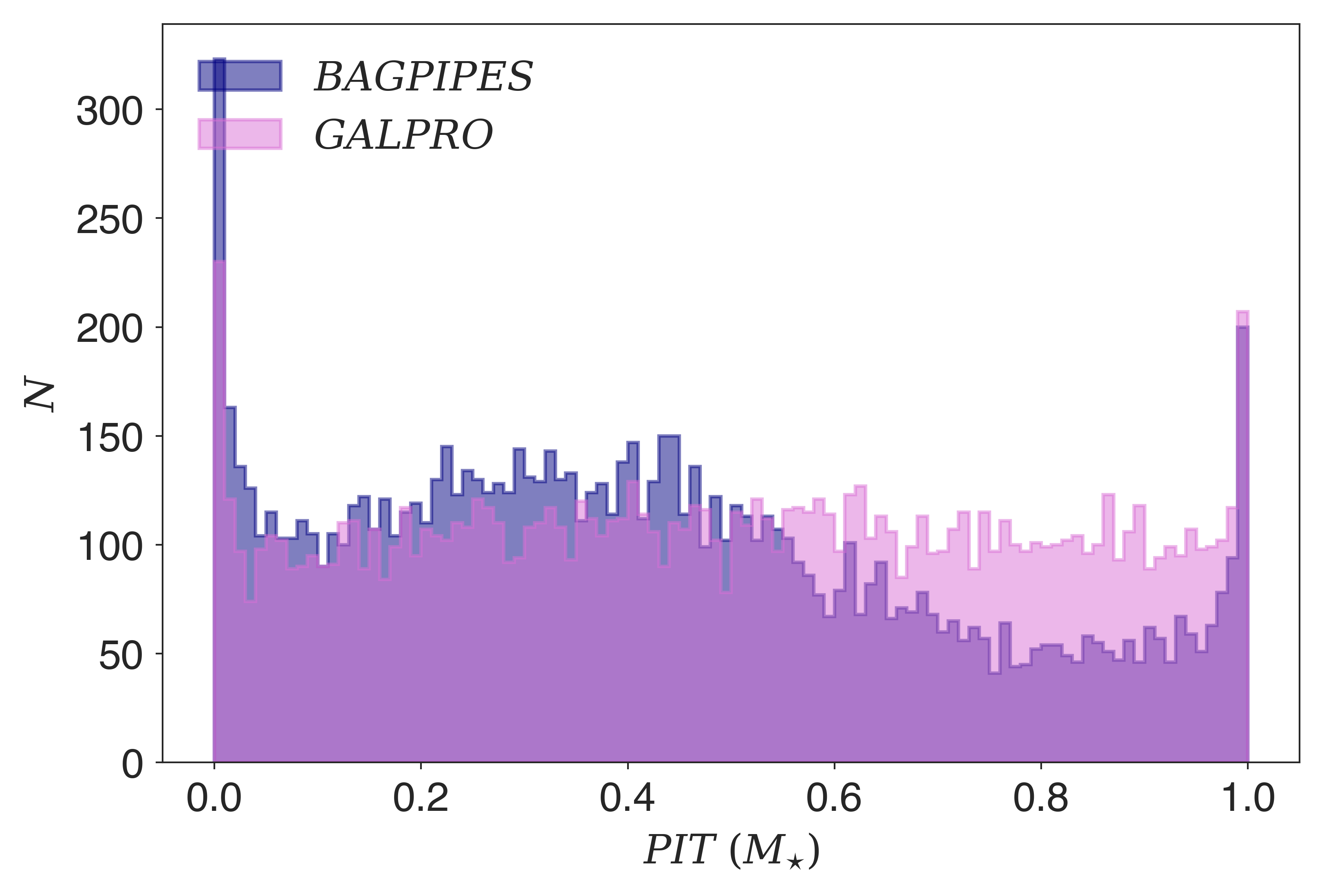}
  \end{subfigure}
  \begin{subfigure}[h!]{0.46\textwidth}
    \includegraphics[width=\textwidth]{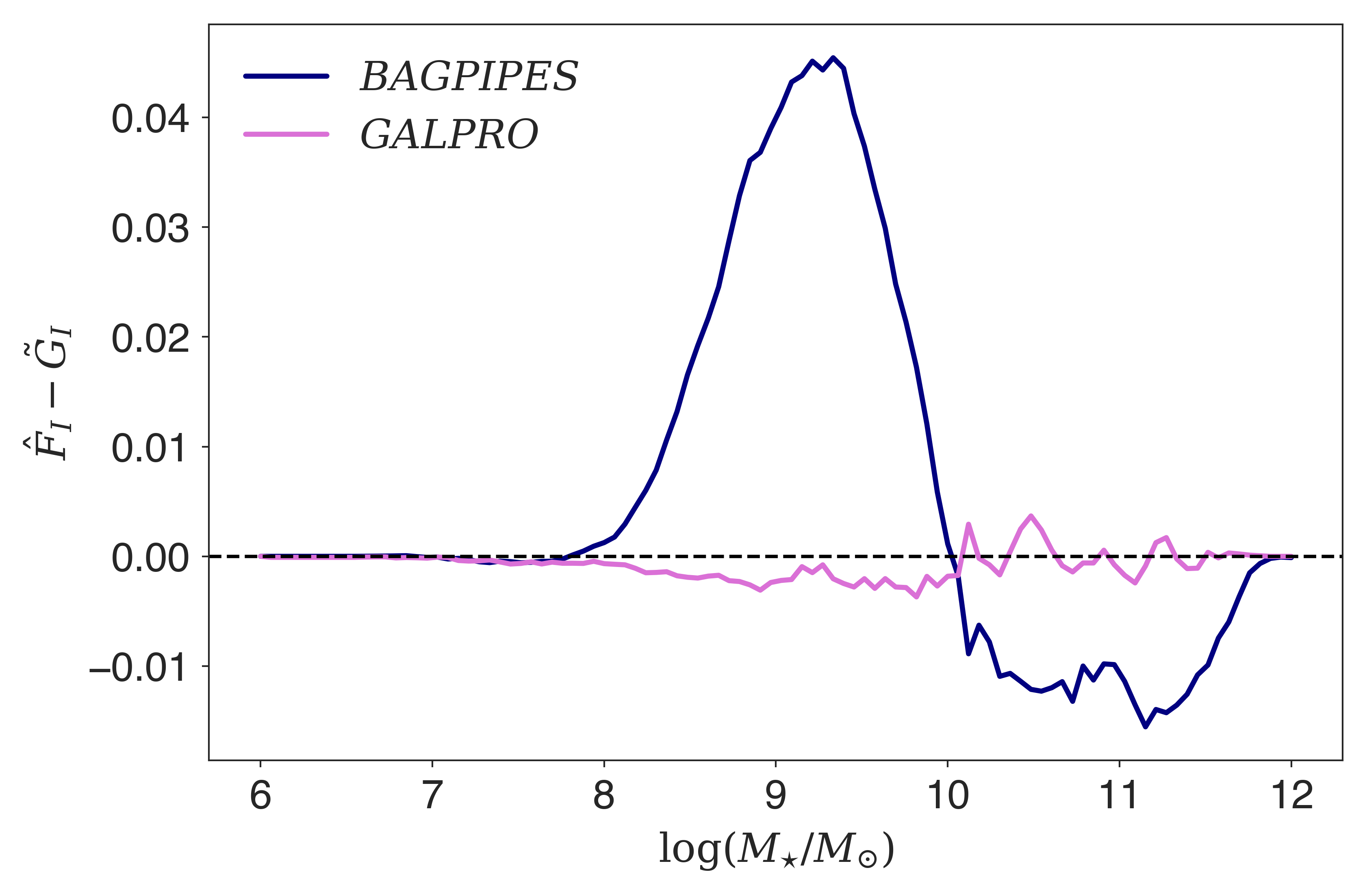}
  \end{subfigure}
  \begin{subfigure}[h!]{0.46\textwidth}
    \includegraphics[width=\textwidth]{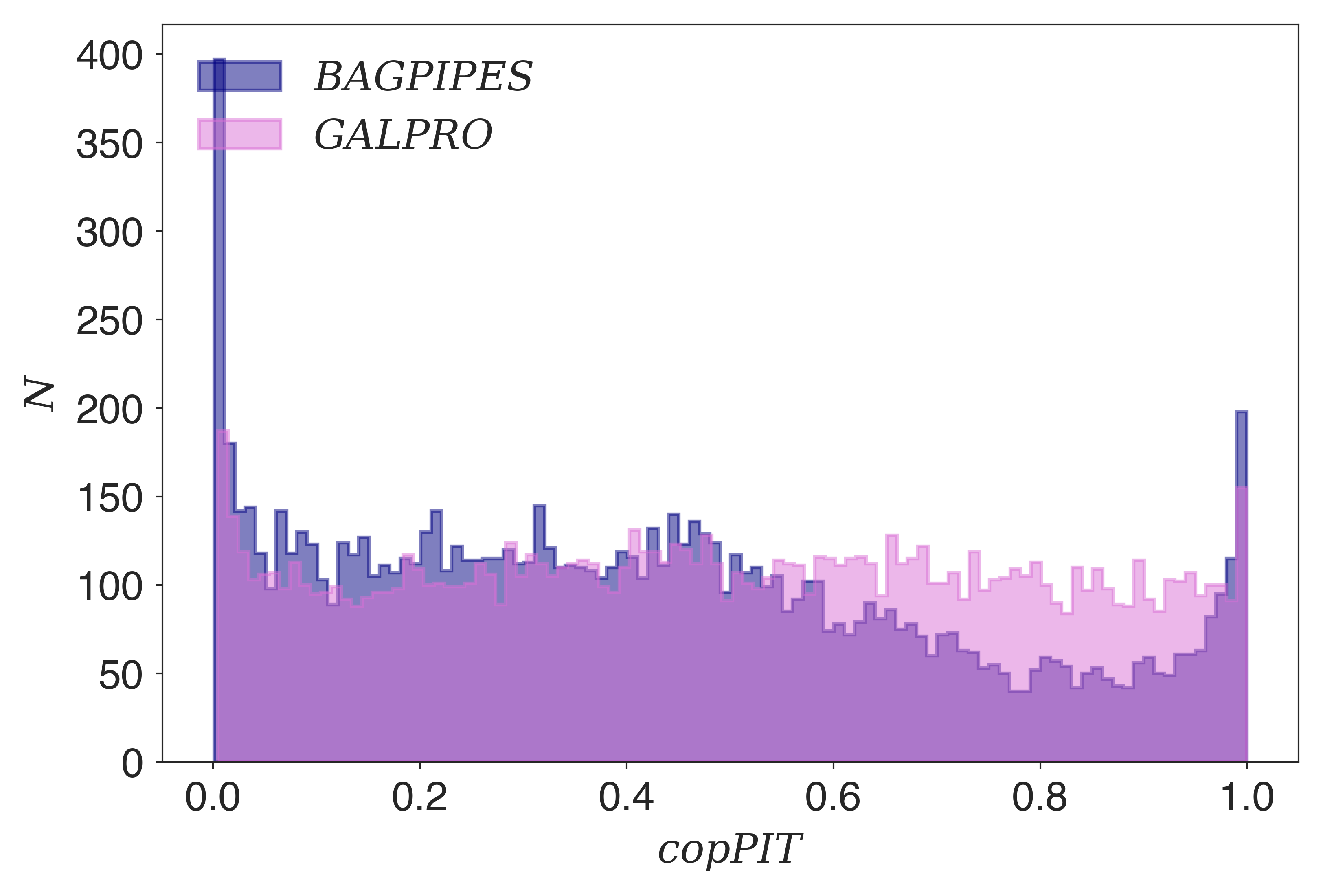}
  \end{subfigure}
  \begin{subfigure}[h!]{0.46\textwidth}
    \includegraphics[width=\textwidth]{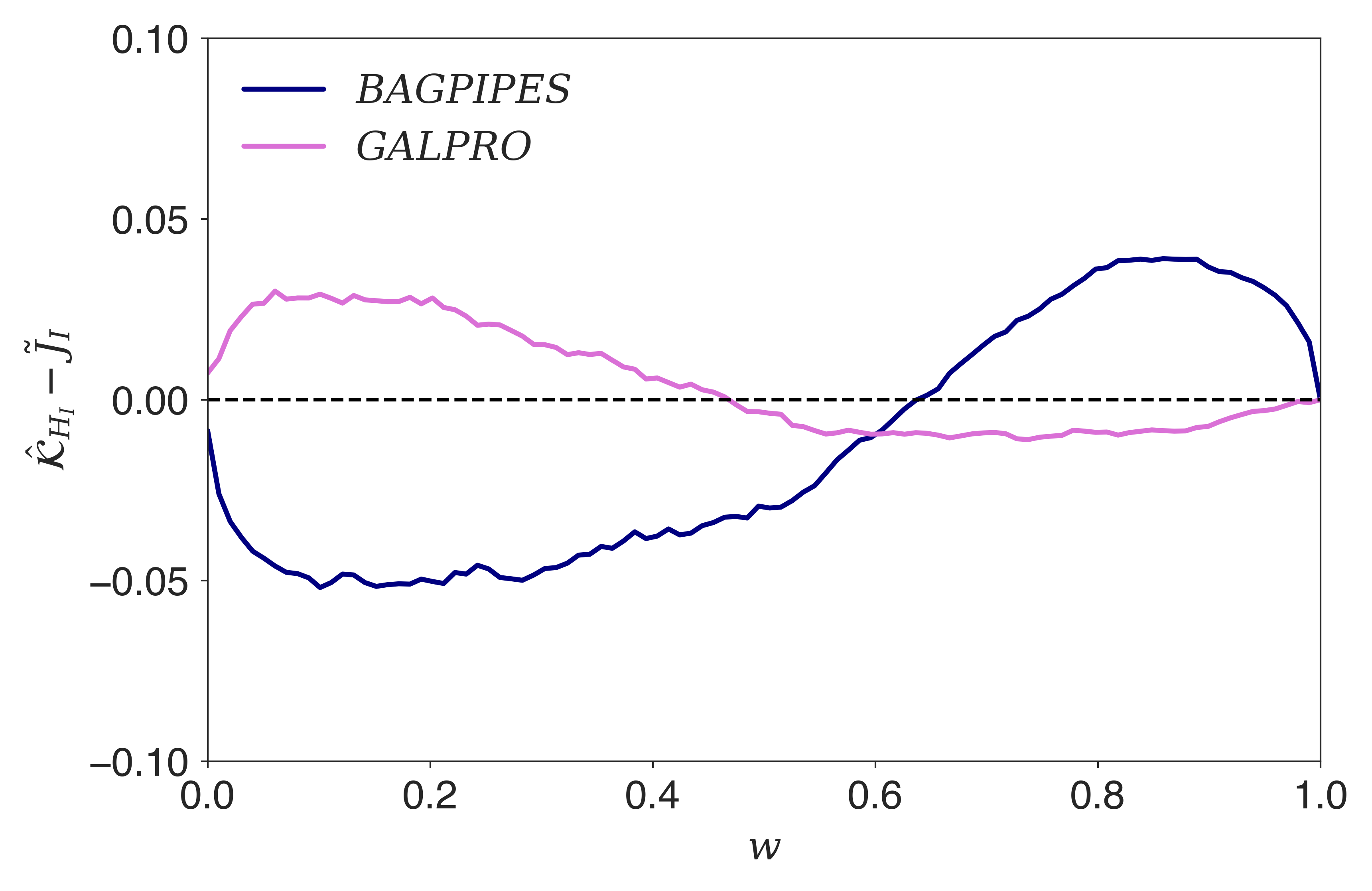}
  \end{subfigure}
  \caption{Comparison diagnostic plots for benchmarking the performance of \texttt{GALPRO} on test galaxies in the DF data set against that of \texttt{BAGPIPES} on a comparable data set, which is composed of the same galaxies but with Subaru photometry in four bands ($V$, $r$, $i+$, and $z++$) from the COSMOS2015 catalogue. The marginal and joint PDFs of redshift and stellar mass produced by \texttt{BAGPIPES} are validated using a `truth' catalogue constructed by running \texttt{BAGPIPES} on photometry in 22 COSMOS bands listed in Tab. \ref{tab:cosmos_bands}.}
  \label{fig:bagpipes}
\end{figure*}

\section{Template-Fitting Comparison}
\label{section:bagpipes}
The different diagnostic plots and the metrics we utilise to validate the marginal and joint PDFs produced by our RF models are difficult to fully appreciate without familiar context. Consequently, we utilise Bayesian Analysis of Galaxies for Physical Inference and Parameter EStimation, or \texttt{BAGPIPES} \citep{bagpipes} to benchmark our results. \texttt{BAGPIPES} is a \texttt{Python} package that uses \texttt{MultiNest} (\citealt{feroz_2008}, \citealt{feroz_2009, feroz_2013}) nested sampling algorithm to model the emission from galaxies and to fit these models to any combinations of spectroscopic and photometric data in order to output multivariate posteriors distributions of parameters such as redshift and stellar mass, hence making it ideal for comparison.

The photometry in the COSMOS2015 and DES Y3 DF catalogues have been calibrated independently of one another. So although we can expect them to be broadly consistent, it is possible that small differences in absolute calibration between the two remain. Even minor offsets in the calibration baseline may have a significant impact on the stellar mass posterior PDFs produced using \texttt{BAGPIPES} with respect to COSMOS2015, and perhaps also some subtle effects in redshift. Accordingly, validation of the PDFs using the point predictions in the catalogue would not be appropriate. To solve this dilemma, we run \texttt{BAGPIPES} on Subaru $V$, $r$, $i+$, and $z++$ bands' photometry from the catalogue in place of the DES DF $griz$ bands. We specifically choose these bands in order to imitate the DES bands as far as possible and therefore allow for an adequate comparison between the template-fitting method and our ML-based method. Although this does not match exactly the degradation in the information provided to the RF, it is nevertheless very similar as we measure PDFs using four optical bands instead of the 30-plus bands available in the catalogue. Importantly, however, we avoid introducing any possible systematic effects that could arise from inter-dataset calibration differences.

The model templates used by \cite{cosmos2015} cannot be exactly reproduced in \texttt{BAGPIPES}. It is important for the validity of our comparison that the four-band PDFs and the truth values are constructed under the same set of model assumptions. Therefore, we produce a new set of truth values using the 22
COSMOS bands (including the four aforementioned) listed in Tab. \ref{tab:cosmos_bands}. In both the 4-band and 22-band runs, we employ the same physical information about the model as outlined in Tab. \ref{tab:priors}. These choices were made to closely mimic the set-up adopted by \cite{cosmos2015} to compute the redshifts and stellar masses in the COSMOS2015 catalogue, so that we can make a fair comparison. There are, however, slight differences that we cannot negate, and as such, a direct comparison is not possible. Nevertheless, they are mostly
similar, and the aggregate metric results should be comparable. We compute total COSMOS flux and flux errors from those measured in a 3 arcsec diameter aperture, and correct for photometric and systematic offsets, and foreground galaxy extinction, before initiating the runs. We define the true values of redshift and stellar mass from the 22-band run to be the mean predictions for each galaxy. Finally, we extract marginal and joint PDFs of redshift and stellar mass from the four-band run and validate them using these new ‘truth’ values. We utilize a total of 14 nodes for both runs, with each node consisting of 12 Xeon X5660 cores and 16GB of random-access memory. The runs take approximately $900$ and $1,400$ h to generate PDFs for $10,699$ galaxies, respectively. Naturally, we only run \texttt{BAGPIPES} on test galaxies in the DF data set. 

Template-fitting with four bands is known to be difficult due to degeneracies in the parameter space \citep[see][for a review]{renzini}. To compensate, authors sometimes restrict the parameter space, for example, by neglecting dust extinction to improve results \citep[e.g.][]{capozzi}, and this amounts to a hard prior in the galaxy population. By design, RF includes an implicit prior built from the training data. We approximate the effect of this prior by applying a 2D population prior formed from the redshifts and stellar masses in the `truth' catalogue to the PDFs estimated by \texttt{BAGPIPES} using the four-band photometry. To apply the prior, we fit a kernel density estimate (KDE) to the `true' redshifts and stellar masses. We use $1\%$ of the total number of point predictions to fit this prior, and this equates to $\sim 200,000$ data points. Next, we compute
the prior probability density at each redshift--stellar mass sample point output by the \texttt{BAGPIPES} nested sampling (with four-band photometry). We produce a smoothed posterior of these points, weighted by the prior probability, via another KDE. Finally, we draw $1000$ importance samples from this smoothed posterior. We repeat this process for all the galaxies. 

We explored the possibility of applying a full 6D prior because, in principle, it should further improve the results. However, doing so caused a large number of galaxies to become catastrophic outliers. It is beyond the scope of this work to go through the painstaking process of carefully optimizing a high-dimensional prior, as we simply want a comparison that assists the reader’s intuition in interpreting the result from our RF models. Nevertheless, we still had a considerable percentage ($6-7\%$) of catastrophic outliers even with our 2D prior. These outliers can skew the performance in terms of the metrics we have chosen and can often be treated separately in scientific analyses. Hence, we remove these outliers and then perform the different calibration checks to better gauge the performance of the population at large. 

Fig. \ref{fig:bagpipes} shows the PIT and the copPIT distributions alongside the marginal and Kendall calibration plots from the analysis, and for comparison, they are overlaid with results from the DES-DF model, labelled as \texttt{GALPRO}. The PIT distributions are not uniform and indicate biased marginal PDFs for the galaxy population, and this correlates well with the marginal calibration plots which have large fluctuations about the zero line. Nevertheless, the marginal redshift PIT distribution is competitive with template-fitting approaches used in code comparison works, e.g. \citet[Fig. 2]{schmidt} and \citet[Fig. 7]{desprez}. However, these studies use deeper data than in this work. Unsurprisingly, a small number of joint PDFs are also biased as reflected by the non-uniform copPIT distribution. Despite the biased PDFs, \texttt{BAGPIPES} does manage to capture the dependence structure between redshift and stellar mass on a similar level to that achieved by the RF. On the whole, RF outperforms \texttt{BAGPIPES} on the metrics we have considered in our analysis. Having said that, it should be possible for \texttt{BAGPIPES} to match the performance of the RF through judicious use of priors and great care in photometric calibration. A great advantage of the RF is that the large effort that would be required to do so is not necessary. An implicit prior is automatically applied, transferring information from the rich training data set to our target data.

To summarize, we have benchmarked the performance of \texttt{GALPRO} against \texttt{BAGPIPES}, and by doing so, we have been able to place our results into context. We have found that our ML-based method performs better in every aspect compared to a template-fitting method that employs a fairly standard set-up. Thus, we have confidence that our models are producing valid marginal and joint posterior probability distributions, based on the different calibration modes and metrics we have employed in our analysis.

\section{Conclusions}
\label{section:conclusions}
The emergence of template-fitting methods with the capability of generating multivariate PDFs of redshift and physical properties of galaxies represents a paradigm shift. These PDFs account for potential correlations between different galaxy properties and fully characterize uncertainties associated with point estimates of the quantities. However, with their potential benefits, comes the task of generating them quickly, which is difficult given their complexity. For example, the template-fitting code \texttt{BAGPIPES} takes a few minutes to fit each galaxy. While this may not seem significant, the amount of time required to generate them for hundreds of thousands of galaxies, let alone the billions that will be observed with the upcoming photometric surveys such as the LSST and \emph{Euclid}, quickly becomes impractical. Coupled with the difficulty of storing such PDFs, a solution that enables on-the-fly production at speed is greatly desirable.

In this work, we tackle the problem by using an ML-based approach. We introduce a novel method based on the RF algorithm to generate joint PDFs. As an example, we generate PDFs for the probability space in redshift and stellar mass, as they are two of the most important to accurately predict. Our method can be generalized to extract n-dimensional PDFs. However, we focus on this specific two-dimensional space as it is easy to visualize and exhibits well-known correlations between the properties.

To demonstrate the method, we train two RF models to produce joint PDFs of galaxies in the DES DF and the main WF DES survey, respectively. We separately combine the COSMOS2015 catalogue, with the DES Y3 DF and the Y3 Balrog to construct the necessary data sets, which contain $53,941$ and $393,276$ galaxies, respectively. From the trained models, we extract point estimates, marginal and joint PDFs of $10,699$ test galaxies. We then proceed to determine the validity of both sets of PDFs, and for this, we utilise the notions of probabilistic copula calibration and Kendall calibration to validate the joint PDFs and their univariate counterparts to validate the marginals. We highlight in particular the advantage of incorporating realistic photometric errors into the RF has on Kendall calibration. We benchmark our results against those achieved by \texttt{BAGPIPES}, adopting a basic set-up and simple population-derived prior in redshift and stellar mass, to provide some context to the metric values and guide our intuition. We find that our ML-based method is producing valid PDFs with only small calibration errors, and performs at a superior level on every metric we consider in our analysis compared to \texttt{BAGPIPES}. Despite the success of our method, template-fitting approaches such as \texttt{BAGPIPES} undoubtedly still have a vital role to play in building the training samples for ML-based codes. 

To conclude, joint redshift--stellar mass PDFs have many potential science applications from determining the evolution of the SMF, to constraining the SHMR. Consequently, we have developed \texttt{GALPRO}, a highly intuitive and efficient \texttt{Python} package for rapidly generating n-dimensional PDFs on-the-fly, thus solving the potential issue of storage. We have trained and tested our RF models using \texttt{GALPRO} on a 13" Macbook Pro (2.4 GHz Intel Core i5, 16GB LPDDR3) and found that, at best, it takes on average a few milliseconds to generate a PDF. Thus, \texttt{GALPRO} can potentially offer a $100,000 x$ reduction in run time compared to packages based on template-fitting methods, making it ideal for the impending era of `big data'. Of course, one must ensure that the training data set is representative and suitable for their scientific analysis to fully reap the benefits of \texttt{GALPRO}.

\section*{Acknowledgements}
SM was supported by the STFC UCL Centre for
Doctoral Training in Data Intensive Science (grant ST/P006736/1).
OL acknowledges support from a European Research Council Advanced grant TESTDE FP7/291329 and an STFC Consolidated grants ST/M001334/1 and ST/R000476/1.
AFLB acknowledges ERC Advanced grant 695671 'QUENCH' and support from the UK STFC.

Funding for the DES Projects has been provided by the U.S. Department of Energy, the U.S. National Science Foundation, the Ministry of Science and Education of Spain, 
the Science and Technology Facilities Council of the United Kingdom, the Higher Education Funding Council for England, the National Center for Supercomputing 
Applications at the University of Illinois at Urbana-Champaign, the Kavli Institute of Cosmological Physics at the University of Chicago, 
the Center for Cosmology and Astro-Particle Physics at the Ohio State University,
the Mitchell Institute for Fundamental Physics and Astronomy at Texas A\&M University, Financiadora de Estudos e Projetos, 
Funda{\c c}{\~a}o Carlos Chagas Filho de Amparo {\`a} Pesquisa do Estado do Rio de Janeiro, Conselho Nacional de Desenvolvimento Cient{\'i}fico e Tecnol{\'o}gico and 
the Minist{\'e}rio da Ci{\^e}ncia, Tecnologia e Inova{\c c}{\~a}o, the Deutsche Forschungsgemeinschaft and the Collaborating Institutions in the DES. 

The Collaborating Institutions are Argonne National Laboratory, the University of California at Santa Cruz, the University of Cambridge, Centro de Investigaciones Energ{\'e}ticas, 
Medioambientales y Tecnol{\'o}gicas-Madrid, the University of Chicago, University College London, the DES-Brazil Consortium, the University of Edinburgh, 
the Eidgen{\"o}ssische Technische Hochschule (ETH) Z{\"u}rich, 
Fermi National Accelerator Laboratory, the University of Illinois at Urbana-Champaign, the Institut de Ci{\`e}ncies de l'Espai (IEEC/CSIC), 
the Institut de F{\'i}sica d'Altes Energies, Lawrence Berkeley National Laboratory, the Ludwig-Maximilians Universit{\"a}t M{\"u}nchen and the associated Excellence Cluster Universe, 
the University of Michigan, NFS's NOIRLab, the University of Nottingham, The Ohio State University, the University of Pennsylvania, the University of Portsmouth, 
SLAC National Accelerator Laboratory, Stanford University, the University of Sussex, Texas A\&M University, and the OzDES Membership Consortium.

Based in part on observations at Cerro Tololo Inter-American Observatory at NSF’s NOIRLab (NOIRLab Prop. ID 2012B-0001; PI: J. Frieman), which is managed by the Association of Universities for Research in Astronomy under a cooperative agreement with the National Science Foundation.

The DES data management system is supported by the National Science Foundation under grants AST-1138766 and AST-1536171.
The DES participants from Spanish institutions are partially supported by MICINN under grants ESP2017-89838, PGC2018-094773, PGC2018-102021, SEV-2016-0588, SEV-2016-0597, and MDM-2015-0509, some of which include ERDF funds from the European Union. IFAE is partially funded by the CERCA program of the Generalitat de Catalunya.
Research leading to these results has received funding from the European Research
Council under the European Union's Seventh Framework Program (FP7/2007-2013) including ERC grants 240672, 291329, and 306478.
We  acknowledge support from the Brazilian Instituto Nacional de Ci\^encia
e Tecnologia (INCT) do e-Universo (CNPq grant 465376/2014-2).

This manuscript has been authored by Fermi Research Alliance, LLC under Contract No. DE-AC02-07CH11359 with the U.S. Department of Energy, Office of Science, and Office of High Energy Physics.

\section*{Data Availability}
The data underlying this article were produced by the DES and the COSMOS. The DES Y3 DF and DES Y3 Balrog catalogues are expected to be available to the public in 2021 and will be hosted at \url{https://des.ncsa.illinois.edu/releases/dr2}. The COSMOS2015 catalogue can be accessed at \url{https://ftp.iap.fr/pub/from_users/hjmcc/COSMOS2015/}. The derived DF and WF data sets will be shared on reasonable request to the corresponding author. The code used to perform all the analysis in this paper and an example data set is available at \url{https://github.com/smucesh/galpro/}.




\bibliographystyle{mnras_2author.bst}
\bibliography{bibliography} 




\section*{Affiliations}
{
\noindent$^{1}$ Department of Physics \& Astronomy, University College London, Gower Street, London, WC1E 6BT, UK\\
$^{2}$ Department of Astronomy, University of Geneva, ch. d’Ecogia 16, CH-1290 Versoix, Switzerland\\
$^{3}$ Fermi National Accelerator Laboratory, P. O. Box 500, Batavia, IL 60510, USA\\
$^{4}$ Kavli Institute for Cosmological Physics, University of Chicago, Chicago, IL 60637, USA\\
$^{5}$ Kavli Institute for Cosmology, University of Cambridge, Madingley Road, Cambridge CB3 0HA, UK\\
$^{6}$ Cavendish Laboratory –- Astrophysics Group, University of Cambridge, 19 JJ Thomson Avenue, Cambridge CB3 0HE, UK\\
$^{7}$ Argonne National Laboratory, 9700 South Cass Avenue, Lemont, IL 60439, USA\\
$^{8}$ Kavli Institute for Particle Astrophysics \& Cosmology, P. O. Box 2450, Stanford University, Stanford, CA 94305, USA\\
$^{9}$ Department of Physics, 2320 Chamberlin Hall, University of Wisconsin-Madison, 1150 University Avenue Madison, WI  53706-1390\\
$^{10}$ Department of Physics and Astronomy, University of Pennsylvania, Philadelphia, PA 19104, USA\\
$^{11}$ Department of Astrophysics Research, Instituto de Astrofisica de Canarias, E-38205 La Laguna, Tenerife, Spain\\
$^{12}$ Dpto. Astrofísica, Universidad de La Laguna, E-38206 La Laguna, Tenerife, Spain\\
$^{13}$ Department of Astronomy, University of Illinois at Urbana-Champaign, 1002 W. Green Street, Urbana, IL 61801, USA\\
$^{14}$ National Center for Supercomputing Applications, University of Illnois at Urbana-Champaign, 1205 West Clark St, Urbana, IL 61801, USA\\
$^{15}$ Center for Cosmology and Astro-Particle Physics, The Ohio State University, Columbus, OH 43210, USA\\
$^{16}$ Santa Cruz Institute for Particle Physics, University of California, Santa Cruz, CA 95064, USA\\
$^{17}$ Department of Physics, Stanford University, 382 Via Pueblo Mall, Stanford, CA 94305, USA\\
$^{18}$ SLAC National Accelerator Laboratory, Menlo Park, CA 94025, USA\\
$^{19}$ Jodrell Bank Center for Astrophysics, School of Physics and Astronomy, University of Manchester, Oxford Road, Manchester, M13 9PL, UK\\
$^{20}$ Jet Propulsion Laboratory, California Institute of Technology, 4800 Oak Grove Dr., Pasadena, CA 91109, USA\\
$^{21}$ Centro de Investigaciones Energ\'eticas, Medioambientales y Tecnol\'ogicas (CIEMAT), 40, 28040 Madrid, Spain\\
$^{22}$ Brookhaven National Laboratory, Bldg 510, Upton, NY 11973, USA\\
$^{23}$ Departamento de F\'isica Matem\'atica, Instituto de F\'isica, Universidade de S\~ao Paulo, CP 66318, S\~ao Paulo, SP 05314-970, Brazil\\
$^{24}$ Laborat\'orio Interinstitucional de e-Astronomia -- LIneA, Rua Gal. Jos\'e Cristino 77, Rio de Janeiro, RJ - 20921-400, Brazil\\
$^{25}$ Institute of Cosmology and Gravitation, University of Portsmouth, Portsmouth PO1 3FX, UK\\
$^{26}$ CNRS, UMR 7095, Institut d'Astrophysique de Paris, F-75014 Paris, France\\
$^{27}$ Sorbonne Universit\'es, UPMC Univ Paris 06, UMR 7095, Institut d'Astrophysique de Paris, F-75014 Paris, France\\
$^{28}$ Department of Physics and Astronomy, Pevensey Building, University of Sussex, Brighton BN1 9QH, UK\\
$^{29}$ Institut de F\'{\i}sica d'Altes Energies (IFAE), The Barcelona Institute of Science and Technology, Campus UAB, E-08193 Bellaterra (Barcelona), Spain\\
$^{30}$ Institut d'Estudis Espacials de Catalunya (IEEC), E-08034 Barcelona, Spain\\
$^{31}$ Astrophysics \& Planetary Sciences, Institute of Space Sciences (ICE, CSIC),  Campus UAB, Carrer de Can Magrans, s/n,  E-08193 Barcelona, Spain\\
$^{32}$ School of Physics and Astronomy, University of Nottingham, Nottingham NG7 2RD, UK\\
$^{33}$ INAF-Osservatorio Astronomico di Trieste, via G. B. Tiepolo 11, I-34143 Trieste, Italy\\
$^{34}$ Institute for Fundamental Physics of the Universe, Via Beirut 2, I-34014 Trieste, Italy\\
$^{35}$ Observat\'orio Nacional, Rua Gal. Jos\'e Cristino 77, Rio de Janeiro, RJ - 20921-400, Brazil\\
$^{36}$ Department of Physics, University of Michigan, Ann Arbor, MI 48109, USA\\
$^{37}$ Department of Physics, IIT Hyderabad, Kandi, Telangana 502285, India\\
$^{38}$ Department of Astronomy and Astrophysics, University of Chicago, Chicago, IL 60637, USA\\
$^{39}$ Department of Astronomy, University of Michigan, Ann Arbor, MI 48109, USA\\
$^{40}$ Institute of Theoretical Astrophysics, University of Oslo. P.O. Box 1029 Blindern, NO-0315 Oslo, Norway\\
$^{41}$ Instituto de Fisica Teorica UAM/CSIC, Universidad Autonoma de Madrid, E-28049 Madrid, Spain\\
$^{42}$ School of Mathematics and Physics, University of Queensland,  Brisbane, QLD 4072, Australia\\
$^{43}$ Department of Physics, The Ohio State University, Columbus, OH 43210, USA\\
$^{44}$ Center for Astrophysics $\vert$ Harvard \& Smithsonian, 60 Garden Street, Cambridge, MA 02138, USA\\
$^{45}$ Australian Astronomical Optics, Macquarie University, North Ryde, NSW 2113, Australia\\
$^{46}$ Lowell Observatory, 1400 Mars Hill Rd, Flagstaff, AZ 86001, USA\\
$^{47}$ Department of Astrophysical Sciences, Princeton University, Peyton Hall, Princeton, NJ 08544, USA\\
$^{48}$ Instituci\'o Catalana de Recerca i Estudis Avan\c{c}ats, E-08010 Barcelona, Spain\\
$^{49}$ Institute of Astronomy, University of Cambridge, Madingley Road, Cambridge CB3 0HA, UK\\
$^{50}$ School of Physics and Astronomy, University of Southampton,  Southampton SO17 1BJ, UK\\
$^{51}$ Computer Science and Mathematics Division, Oak Ridge National Laboratory, Oak Ridge, TN 37831\\
$^{52}$ Max Planck Institute for Extraterrestrial Physics, Giessenbachstrasse, 85748 Garching, Germany\\
$^{53}$ Fakult\"at f\"ur Physik, Universit\"ats-Sternwarte, Ludwig-Maximilians Universit\"at M\"unchen, Scheinerstr 1, D-81679 M\"unchen, Germany\\
}

\bsp	
\label{lastpage}
\end{document}